\documentclass[
 reprint,
 preprintnumbers,
 amsmath,
 amssymb,
 aps,
 prstab,
 floatfix,
]{revtex4-1}

\usepackage{graphicx}			% Graphics
\usepackage{verbatim}			% Graphics
\usepackage{wasysym}			% Bold PI
\usepackage{mathtools}			% Curved arrow
\usepackage[makeroom]{cancel}		% Crossed math
\usepackage{dcolumn}			% Align table columns on decimal point
\usepackage{hyperref}			% Add hypertext capabilities
\usepackage{color}			% DO NOT REMOVE!!!

\newcommand{\pd}{\partial}				% Partial derivative
\newcommand{\dd}{\mathrm{d}}				% General derivative

\newcommand{\Nb}{N_{\text{b}}}		% N_b number of particles in bunch
\newcommand{\Qb}{ Q_{\beta}}		% Q_b betatron tune
\newcommand{\Qs}{ Q_{\text{s}}}		% Q_s syncrotron tune
\newcommand{\Qef}{Q_{\text{eff}}}	% Q_eff tune shift
\newcommand{\Qmx}{Q_{\text{max}}}	% Q_max SC tune shift
\newcommand{\Qw}{ Q_{\text{W}}}		% Q_w wake-driven coherent tune shift

\newcommand{\Qsc}{\Delta Q_\mathrm{sc}}
\newcommand{\Qx }{\Delta Q_k          }
\newcommand{\qx }{\Delta q_k          }
\newcommand{\tb }{       \tau_\mathrm{b }}
\newcommand{\cb }{       \sigma_\mathrm{b }}
\newcommand{\vb }{\mathrm{v }_\mathrm{b }}

\newcommand{\vv  }{         \mathrm{v }}

\newcommand{\Yk}{\mathcal{Y}_k}				% modes Y_k

\newcommand{\hG}{\widehat{\mathrm{G}}}			% matrix G
\newcommand{\hW}{\widehat{\mathrm{W}}}			% matrix W
\newcommand{\hD}{\widehat{\mathrm{D}}}			% matrix D
\newcommand{\mG}{\widehat{\mathrm{G}}_{lm}}		% matix element П_lm
\newcommand{\mW}{\widehat{\mathrm{W}}_{lm}}		% matix element Ц_lm
\newcommand{\mD}{\widehat{\mathrm{D}}_{lm}}		% matix element D_lm
\newcommand{\mL}{\widehat{\Lambda   }_{lm}}		% matix element Л_lm

\newcommand{\ds}{\displaystyle}			% Skew   mltpl strength

\begin{document}

%===============================================================================
\title{TMCI and Space Charge}
\author{T.~Zolkin}
\email{zolkin@fnal.gov}
\affiliation{Fermilab, PO Box 500, Batavia, IL 60510-5011}
%\thanks{Operated by Fermi Research Alliance, LLC under Contract
%No.~De-AC02-07CH11359 with the U.S. Department of Energy.}
\author{A.~Burov}
\affiliation{Fermilab, PO Box 500, Batavia, IL 60510-5011}
\author{B.~Pandey}
\affiliation{The University of Chicago, 5801 S Ellis Ave, Chicago, IL 60637}
\date{\today}
%===============================================================================

%===============================================================================
\begin{abstract}
Transverse mode-coupling instability (TMCI) is known to limit bunch intensity. 
Since space charge (SC) changes coherent spectra, it affects the TMCI threshold.
Generally, there are only two types of TMCI with respect to SC: the vanishing 
type and the strong space charge (SSC) type.
For the former, the threshold value of the wake tune shift is asymptotically 
proportional to the SC tune shift, as it was  first observed twenty years ago by 
M. Blaskiewicz for exponential wakes.
For the latter, the threshold value of the wake tune shift is asymptotically 
inversely proportional to the SC, as it was shown by one of the authors.
In the presented studies of various wakes, potential wells, and bunch
distributions, the second type of instability was always observed for cosine 
wakes;
it was also seen for the sine wakes in the case of a bunch within a square 
potential well.
The vanishing TMCI was observed for all other wakes and distributions we 
discuss in this paper: always for the negative wakes, and always, except the 
cosine wake, for parabolic potential wells.
At the end of this paper, we consider high-frequency broadband wake, suggested
as a model impedance for CERN SPS ring.
As expected, TMCI is of the vanishing type in this case.
Thus, SPS Q26 instability, observed at strong SC almost with the same bunch 
parameters as it would be observed without SC, cannot be TMCI.
\end{abstract}
%===============================================================================

% PACS, the Physics and Astronomy Classification Scheme.
% Use showkeys class option if keyword
                             % display desired

%===============================================================================
\maketitle
%===============================================================================

%==============================================================================%
%==============================================================================%
%==============================================================================%
\section{Introduction}

%------------------------------------------------------------------------------%
The problems of coherent beam stability are known to be hard when the beam
space charge (SC) has to be taken into account, which is necessary for low- and 
medium-energy hadron rings.
So far the only universally accurate method available is macroparticle 
tracking;
since the number of needed macroparticles per bunch may be up to a billion for 
reliable
results, this method is very expensive in terms of CPU time.
This is why analytical models are valuable: although each of them
has limitations, they help build general understanding, providing
important results within their areas of validity.
Since these areas, as well as the models' accuracy,  are not always clear, 
the results of available analytical models should be compared whenever possible.

%------------------------------------------------------------------------------%
The transverse mode coupling instability (TMCI), also known as the strong 
head-tail instability, is one of the main intensity limitations of bunched 
beams in circular machines~\cite{chao1993physics}.
The specifics and even existence of this instability for beams with strong space
charge, i.e. when the SC tune shift exceeds the synchrotron tune, remained 
unclear for a rather long time.
A breakthrough publication in this direction, authored by 
M.~Blaskiewicz, appeared about twenty yeas ago~\cite{blaskiewicz1998fast}.
A model of an AirBag bunch within a Square potential well (addressed here as 
the ABS model) was presented and analyzed there with the exponential wake and 
an 
arbitrary 
ratio of the SC tune shift to the synchrotron tune, or 
{\it the space charge parameter}.
A method to extend this model to an arbitrary sum of exponential/oscillating 
wakes was also described.
It was shown that for a wide range of the wake decay rate and the SC parameter
the instability threshold grows with the latter.
A qualitative explanation of why SC may work this way was suggested the next
year~\cite{ng1999stability}.
In 2009, a theory of head-tail instabilities with strong space charge 
(SSC) was presented in Refs.~\cite{burov2009head,balbekov2009transverse}.
One of the authors of this paper speculated then that dependence of the TMCI 
threshold on the SC parameter might be non-monotonic, meaning that the 
threshold 
might
start to decrease after a certain value of the SC tune shift.
It was demonstrated that for the cosine wake this statement is true, but 
whether 
it 
is true for sign-constant wakes remained unclear.
About a year ago, V.~Balbekov confirmed his agreement \cite{balbekov2016tmci} 
with this hypothesis for the constant wake and the {\it boxcar model} (referred 
to as 
HP$_0$ distribution in this paper).
Soon after that, however, he withdrew this result, claiming that for both ABS 
and boxcar models the threshold grows monotonically with the space charge 
parameter, showing no hint that this trend would change at a higher SC 
\cite{balbekov2016dependence,balbekov2016tmci,balbekov2017transverse,
PhysRevAccelBeams.20.114401}.
It has been also shown \cite{balbekov2017transverse} that for the cosine 
wake with the phase advance not exceeding $\pi$, the TMCI threshold 
monotonically increases with the SC parameter, similarly to the 
negative wakes.

%------------------------------------------------------------------------------%
In this paper, we are trying to get a better confidence and wider vision of the 
TMCI threshold versus the SC parameter for various wakes, potential wells and 
bunch longitudinal distributions.
Our paper is a summary of results for several models. 

%------------------------------------------------------------------------------%
We use two analytical approaches: the ABS of M.~Blaskiewicz and the Strong 
Space Charge (SSC) theory of Ref.~\cite{burov2009head}.
While the ABS model can be applied for any SC strength, the SSC theory is 
applicable only when the SC tune shift sufficiently exceeds both the 
synchrotron tune and the coherent tune shift due to the wake.
However, the SSC theory has an advantage in its applicability for any potential 
wells and longitudinal distributions, so the two approaches complement each 
other.
Within the SSC approach, we examine first the square potential well 
with an arbitrary longitudinal distribution, and then the boxcar distribution 
for the 
parabolic potential well.
The former of these we denote as SSCSW (strong space charge, square well), and 
to
the latter we refer as the generalized Hofmann-Pedersen 
distribution of zeroth order, or SSCHP$_0$ model; the naming was suggested
by~\cite{hofmann1979bunches}.

%------------------------------------------------------------------------------%
In addition to SSCHP$_0$, we also consider elliptic arc and parabolic line 
densities 
for the same potential well, or SSCHP$_{1/2}$ and SSCHP$_{1}$ models.
On top of that, TMCI for a Gaussian bunch at the strong space charge 
limit (SSCG) is discussed for various wakes. 

%------------------------------------------------------------------------------%
Recently, the two-particle model for a bunch with SC and a constant wake has 
been suggested in~\cite{chin2016two}.
We think that its main advantage, simplicity, is somewhat excessive for
the specific problem under study, so we leave its possible application outside
the framework of this paper. 

%------------------------------------------------------------------------------%
For the negative exponential wakes our conclusion agrees 
with M.~Blaskiewicz
\cite{blaskiewicz1998fast} and the latest results of V.~Balbekov
\cite{balbekov2017transverse}: SC always elevates the 
TMCI threshold;
when the SC parameter is large, the threshold wake tune shift grows linearly 
with 
it.
This result is also confirmed with all SSC models for the resistive wall wake.

%------------------------------------------------------------------------------%
For the cosine wakes and strong space charge, the situation is different. 
When SC is strong enough, it makes the beam less stable; the threshold is found 
to be 
inversely proportional to the SC parameter, provided that the wake phase 
advance 
is 
sufficiently large, $\omega\,\tb>\pi$.
However, for the same wake with a small or moderate SC parameter, the 
thresholds 
typically show some growth with SC, thus confirming the hypothesis of 
Ref.~\cite{burov2009head} about non-monotonic dependence of the TMCI threshold 
on the SC parameter for oscillating wakes.
When the wake oscillations are not well-pronounced (either the phase advance is 
insufficient, $\omega\,\tb < \pi$, or the wake oscillations are overshadowed by 
their exponential decay), the TMCI vanishes at the SSC limit, as expected.

%------------------------------------------------------------------------------%
For the sine wake at SSC, results for the square and parabolic potential wells 
were found to be qualitatively different.
For the SSCSW model the TMCI behaves in a similar manner as for the cosine 
wake, but every mode coupling is followed by its decoupling at higher wake 
amplitude.
Contrary to that, the modes never couple for the SSCHP$_0$ model; they either 
cross or approach-divert, as if the instability is prohibited for this case.
For more realistic SSCHP$_{1/2,1}$ and SSCG models, we found that there is no 
TMCI either, even without any crossings or approach-diverts. 
Thus, from our numerical results we conclude that TMCI vanishes for the 
parabolic potential well and the sine wake.

%------------------------------------------------------------------------------%
It will be shown in Section~\ref{sec:Models} that there are only two types 
of TMCI with respect to the SC.
The first one, called here {\it vanishing TMCI}, is such that its threshold 
wake tune shift
asymptotically scales as 
the SC tune shift. For this sort of instability, there is an absolute threshold 
of the wake amplitude, inversely proportional to the transverse emittance: 
below 
this threshold, the beam is stable for any number of particles. For the second 
type of TMCI, which we call {\it SSC TMCI}, the wake 
threshold 
asymptotically drops in the inverse proportion to the
SC tune shift. No other type of the asymptotic threshold behavior is possible. 

%------------------------------------------------------------------------------%
%The reason that two, and only two, types of TMCI at strong space charge are 
%possible is in asymptotic nature of the SSC theory:
%it assumes the {\it rigid-slice approximation} based on the idea that 
%transverse
%coherent motion can be described as a displacement of undeformable 
%longitudinal 
%slices.
%When SSC theory is valid (i.e SC tune shift is much greater than $k\,\Qs$ 
%where 
%$k$ is the mode number), the effective SC tune shift only plays a role by 
%means 
%of the coherent tune shift units, $\Qs^2/\Qef(0)$.
%Due to this scaling invariance, if modes do not couple in the SSC case, the 
%TMCI 
%threshold have to grow at least as fast as the SC parameter.
%Unlike, when modes do couple in the SSC approximation, the asymptotic behavior 
%of the threshold is inversely proportional to the SC parameter, due to the 
%reduction of distance between modes in the positive part of spectrum.
%Below, by considering different bunch distributions and wake functions,
%we will try to determine to which situation this or that case belongs.

%==============================================================================%
%==============================================================================%
\subsection{Article structure}

%------------------------------------------------------------------------------%
The paper is structured as follows.
First, Sec.~\ref{sec:Models} summarizes the SSC and ABS main formulas for a 
single bunch at zero chromaticity for the reader's convenience.
Modes for the no-wake case, or the {\it SSC harmonics}, are presented in the
subsection~\ref{sec:SSCTheory} for all strong space charge models.
Subsection~\ref{sec:AB} describes the airbag model including the solution for a
no-wake case.
Two Sections \ref{sec:NegWakes} and \ref{sec:OscWakes} are dedicated to 
the negative (delta-functional, constant, step-function, exponential and 
resistive wall) and oscillating (sine, cosine and resonator) wake functions 
respectively.
In subsection~\ref{sec:ROsc}, dedicated to resonator wakes, we compare SSC
theory with the results of simulations~\cite{blaskiewicz2012comparing}.
The last subsection~\ref{sec:SPS} presents TMCI threshold versus space charge 
parameter for the high-frequency broadband wake suggested for CERN SPS ring, 
using for that the ABS model; these results are compared with simulations of 
Ref.~\cite{quatraro2010effects}.
Appendix~\ref{secAP:Wlm} contains the expressions of the wake matrix elements 
$\mW$
for the SSCHP$_0$ and details of their computation  
for other models.
Appendix~\ref{secAP:AB} provides the reader with details on the exponential and 
trigonometric wakes for the ABS model.
Since some of the results are qualitatively similar to each other, the last 
complementary Appendix~\ref{secAP:Spectra} includes all bunch spectra for SSC 
models not presented into the main text.

%\newpage
%==============================================================================%
%------------------------------------------------------------------------------%
%------------------------------------------------------------------------------%
%------------------------------------------------------------------------------%
%==============================================================================%
\section{\label{sec:Models}Analytical models}

%==============================================================================%
%------------------------------------------------------------------------------%
%==============================================================================%
\subsection{\label{sec:SSCTheory}Strong space charge theory}

%------------------------------------------------------------------------------%
In this subsection, we recall the main formulas of the SSC 
theory~\cite{burov2009head}, which accuracy was verified by multiparticle 
tracking simulations of Ref~\cite{PhysRevSTAB.18.074401}.
Consider a single bunch with a longitudinal distribution function $f(\tau,\vv)$,
where $\tau$ is the position along the bunch, and $\vv$ is the particle 
longitudinal velocity, $\vv=\dd\tau/\dd \theta$.
Both the position $\tau$ and the time $\theta$ are measured in radians.
For zero wake, transverse modes satisfy an ordinary differential equation with
zero-derivative boundary conditions, which composes a standard Sturm-Liouville
(S-L) problem,
\begin{equation}
\label{math:NoWakeMainEq}
\left\{\begin{array}{l}\ds
	\frac{1}{\Qef(\tau)}\frac{\dd}{\dd\tau}
	\left[
		u^2(\tau) \frac{\dd\,Y(\tau)}{\dd \tau}
	\right]
	 + \nu\,Y(\tau)
	= 0,							\\[0.5cm]
	\ds
	\left.\frac{\dd}{\dd\tau}\,Y(\tau) \right|_{\tau = \pm\infty} = 0.
\end{array}\right.
\end{equation}
Solutions of this problem constitute an orthogonal basis, which when normalized
will be referred to as the {\it SSC harmonics} 
$[\nu_k,Y_k(\tau)]$:
\begin{equation}
	\int_\text{SB}\rho(\tau)\,Y_l(\tau)\,Y_m(\tau)\,\dd\tau = \delta_{lm},
\end{equation}
with $\rho(\tau)$ being the normalized line density
\begin{equation}
\label{math:NormCond}
	\rho(\tau) = \int f(\tau,\vv) \,\dd\vv:\qquad
	\int_\text{SB} \rho(\tau)\,\dd\tau = 1.
\end{equation}
The SB integration limit indicates integration along the bunch length of the 
single bunch.
The temperature function $u^2(\tau)$ is the local average of the longitudinal
velocity squared,
\begin{equation}
	u^2(\tau) = \int f(\tau,\vv)\,\vv^2\,\dd \vv
	\,\bigg/\,\rho(\tau).
\end{equation}
$\Qef$ is the {\it effective space charge tune shift} at the given position
along the bunch,
\begin{equation}
	\Qef(\tau) = \Qef(0)\,\frac{\rho(\tau)}{\rho(0)},
\end{equation}
where ``effective'' means the transverse average at the given position $\tau$ 
along the bunch.
If the transverse distribution is Gaussian, the effective space charge tune 
shift is~$\approx 0.52$ of its value at the beam axis. 

%------------------------------------------------------------------------------%
The wake function $W(\tau)$ modifies the collective dynamics as follows:
\begin{equation}
\label{math:Woperator}
	\frac{1}{\Qef(\tau)}\frac{\dd}{\dd\tau}
	\left[
		u^2(\tau) \frac{\dd\,\mathcal{Y}(\tau)}{\dd \tau}
	\right]
	+ \Delta q\,\mathcal{Y}(\tau)
	= \kappa\,\hW\,\mathcal{Y}(\tau), 
\end{equation}
with
\begin{equation}
	\hW\,\mathcal{Y}(\tau) = \int_{\tau}^\infty
	W(\tau-\sigma)\,\rho(\sigma)\,\mathcal{Y}(\sigma)\,
	\dd\sigma,
%\hD\,Y(\tau) = Y(\tau)
%	\int_{\tau}^\infty
%		D(\tau-\sigma)\,\rho(\sigma)\,
%	\dd\,\sigma
\end{equation}
and
% $\zeta = -\xi/\eta$ is the negated ratio of conventional chromaticity, $\xi$,
% and a slippage factor, $\eta = \gamma_\text{t}^{-2}-\gamma^{-2}$.
\begin{equation}
	\kappa = \Nb\,\frac{r_0 R_0}{4\,\pi\,\gamma\,\beta^2\,\Qb}.
\end{equation}
Here $\Nb$ is the number of particles per bunch,
$r_0$ is the particle classical radius,
$R_0=C_0/(2 \pi)$ is the average accelerator ring radius,
$\Qb$ is the bare betatron tune, 
$\gamma$ is the Lorentz factor
and
$\beta$ is the ratio of particle velocity to the speed of light. 
Zero-derivative 
boundary condition, as in Eq.~(\ref{math:NoWakeMainEq}), is assumed. 
Equation~(\ref{math:Woperator}) implies zero chromaticity, which is
always the case in this paper. 

Note that this eigensystem problem has a special scaling with respect to the 
SC. 
If $\{Y, \Delta q \}$ are an eigenfunction and eigenvalue for certain SC tune 
shift $\Qef$ and wake $W$, then for $\alpha$ times larger SC, $\alpha\,\Qef$, 
and $\alpha$ times smaller wake, $W/\alpha$, the eigenvalues scale as $\Delta 
q/\alpha$, while the eigenfunctions $Y$ remain the same. Consequently, the 
instability threshold, if there is any, scales inversely proportional to the SC 
tune shift $\kappa W_{\mathrm{th}} \propto 1/\Qef(0)$. That sort of 
instabilities, 
seen 
within the SSC asymptotic approach, we call {\it SSC TMCI}. When modes of 
Eq.~(\ref{math:Woperator}) never couple, it does not mean that TMCI is 
impossible. Instead, it means that TMCI can only be outside the SSC 
applicability area, when some of wake tune shifts are comparable with the SC 
tune shift of the core particles, $\Qef(0)$. When the SC-asymptotic of the TMCI 
threshold is considered, $\Qef(0)$ exceeds all other tune shifts, which means 
that the TMCI threshold, if it does exist at all, asymptotically has to be  
proportional to the SC tune shift, $\kappa W_{\mathrm{th}} \propto \Qef(0)$. 
This type of TMCI we call {\it vanishing}, since it cannot be seen at SSC. An 
interesting feature of the vanishing type is that in this case there is an {\it 
absolute threshold} of the wake amplitude: if the wake is smaller that this 
threshold, the beam is stable for any number of particles. It is 
straightforward 
to show that the absolute wake threshold is inversely proportional to the 
transverse emittance and average beta-function. Let us stress one more time, 
that except these two types of TMCI no other type is possible, whatever are 
wakes, potential wells and bunch distributions.

%------------------------------------------------------------------------------%
Equation~(\ref{math:Woperator}) can be solved by means of expansion over the 
orthonormal basis of its zero-wake eigenfunctions, or the SSC harmonics,
\begin{equation}
\label{math:EigenDecomp}
	\Yk(\tau) = \sum_{i=0}^{\infty} \mathbf{C}^{(k)}_i Y_i(\tau)
\end{equation}
leads to the eigenvalue problem
\begin{equation}
\label{math:EVProb}
\mathbf{M}\cdot\mathbf{C}^{(k)} = \qx \mathbf{C}^{(k)},
\qquad
\mathbf{M}_{lm} = \nu_l\,\delta_{lm} + \kappa\,\mW,
\end{equation}
with matrix elements of the wake operator being
\begin{equation}
	\mW = \int_{-\infty}^\infty \int_{\tau}^\infty
		W(\tau-\sigma)\,
		\rho(\tau) \, \rho(\sigma) \, Y_l(\tau) \, Y_m(\sigma)
		\,\dd\,\sigma\,\dd\,\tau.
%\mD &=& \int_{-\infty}^\infty \int_{\tau}^\infty
%	\dd\,\sigma\,\dd\,\tau\,D(\tau-\sigma)\,
%		\rho(\tau) \, \rho(\sigma) \, Y_l(\tau) \, Y_m(\tau)\,,	\\
%\mG &=& \mathrm{K}_l(\zeta)\,\mathrm{P}^*_m(\zeta)\,, 
%\quad\text{where}							\\
%\mathrm{P}_k\,\, &=& \int_{-\infty}^\infty\,\dd\,\tau\,
%	P(\tau) \, \rho(\tau) \, Y_k(\tau) \, e^{i\,\zeta\,\tau}\,,	\\
%\mathrm{K}_k\,\, &=& \int_{-\infty}^\infty\,\dd\,\tau\,
%	K(\tau) \, \rho(\tau) \, Y_k(\tau) \, e^{i\,\zeta\,\tau}\,.
\end{equation}
The eigenvalues $\qx$ constitute the bunch spectrum, and once they have been 
determined we look for imaginary values corresponding to coherent growth rates 
of instabilities.
Real parts of $\qx$ correspond to coherent tune shifts and are equal to $\nu_k$ 
for the no-wake case.
In order to study the effect of the wake on the bunch stability, we introduce a
parameter combining the intensity and the wake amplitude, $W_0$,
\begin{equation}
\label{math:chi}
	\chi^* = \frac{\kappa\,W_0\,\Qef(0)}{\Qs^2}.
\end{equation}

%------------------------------------------------------------------------------%
As it was demonstrated by V.~Balbekov \cite{balbekov2017transverse}, the
expansion over the SSC harmonics may have a poor convergence, so that the
numerically obtained instability threshold is very sensitive to the accuracy of 
the matrix elements computation and the number of harmonics taken into account.
In order to distinguish the illusory, purely numerical, threshold from the real
one, two things are required: a large number of basis harmonics and high
accuracy of the computed matrix elements.
To provide that, we considered two different distributions with constant line 
density (SSCSW and SSCHP$_0$ defined below), where the matrix elements can be 
computed analytically, thus minimizing numerical errors up to machine precision 
and justifying the inclusion of large number of modes into analysis.
Analytical expressions for the matrix elements $\mW$ are provided in 
Appendix~\ref{secAP:WlmCONST}.

%------------------------------------------------------------------------------%
In order to compare results of these rather simplified models, some cases were 
considered with a parabolic potential well and smoother bunch distributions, 
including the Gaussian one.
The eigensystems for the first 40 modes were found using
\texttt{Wolfram Mathematica 11.0}.
Then, the matrix elements were calculated using numerical integration
(see Appendix~\ref{secAP:WlmNUMER} for details).
Below we give definitions of all our SSC models and describe SSC harmonics.

%==============================================================================%
%==============================================================================%
\subsubsection{\label{sec:SSCSW}Square potential well}

%------------------------------------------------------------------------------%
Our first SSC model relates to a bunch in the Square potential Well (SSCSW).
In this case the distribution function is factorized
\begin{equation}
	f_\text{sw}(\tau,\vv) =
		\frac{1}{\tb}\,\text{H}\left[ 1-4\,\frac{\tau^2}{\tb^2}\right]
		\mathrm{V}(\vv).
\end{equation}
Below we omit the Heaviside function, assuming that all bunch related functions 
are defined on its length, $|\tau| \leq \tb/2$.
The average square of velocity is position-independent, 
\begin{equation}
	u_\text{sw}^2(\tau) \equiv \vb^2.
\end{equation}

%------------------------------------------------------------------------------%
Then, the equation for the SSC harmonics can be written as
\begin{equation}
	0 = \nu_k\,Y_k^\text{sw}(\tau) +
		\frac{\vb^2}{\Qef(0)}\frac{\dd}{\dd\tau}\left[
		\frac{\dd}{\dd\tau}\,Y_k^\text{sw}(\tau)
	\right].
\end{equation}
With $\tau$ measured in units of $\tb$ and $\nu$ in units of 
$Q_s^2/\Qef(0) \equiv \pi^2\vb^2/[\tb^2\Qef(0)]$, it reduces to
\begin{equation}
\left\{
	\begin{array}{l}\displaystyle
		Y''(\tau) + \pi^2\nu\,Y(\tau) = 0,	\\[0.25cm]
		Y'(\pm 1/2) = 0.
	\end{array}
\right.
\end{equation}
This S-L problem yields the eigenvalues as integers squared,
\begin{equation}
	\nu_k^\text{sw} = \left\{k^2\,|\,k\in\mathbb{Z} \right\} =
	\left\{ 0,1,4,9,25,36,\ldots \right\}.
\end{equation}
The set of eigenfunctions, normalized according to the general SSC harmonics 
orthonormalization rule~(\ref{math:NormCond}), consists of the constant 0-th 
mode, $Y^\text{sw}_0 = 1$, and a sequence of cosine and sine functions
(see Fig.~\ref{fig:SSCHarm})
\begin{equation}
\label{math:YkSW}
Y_k^\text{sw}(\tau) = \sqrt{2-\delta_{0k}}\times\left\{
\begin{array}{lll}
	\displaystyle \cos(\pi\,k\,\tau),	&
	\qquad & k\text{ is even},		\\[0.25cm]
	\displaystyle \sin(\pi\,k\,\tau),	&
	\qquad & k\text{ is odd}.
\end{array}
\right.
\end{equation}

%==============================================================================%
%==============================================================================%
\subsubsection{\label{sec:SSCHP}Hofmann-Pedersen distribution}

%------------------------------------------------------------------------------%
Another SSC model, where matrix elements for some wakes can be analytically 
calculated, is a bunch with a constant line density inside a parabolic potential
well.
Following M.~Blaskiewicz, this distribution was previously referred as a
{\it boxcar}, but since more than one of our models have a boxcar, or constant, 
line density, in order to avoid confusions, we refer to it as the $\text{HP}_0$ 
model, or zeroth Hofmann-Pedersen distribution.
Its phase space density can be written as a function of the Hamiltonian
\begin{equation}
f_{\text{HP}_{0}}(\tau,\vv) =
	\frac{2}{\pi\,\tb\vb}\,
	\left(1-\frac{\mathcal{H}}{\mathcal{H}_b}\right)^{-1/2}
	\text{H}\left[1-\frac{\mathcal{H}}{\mathcal{H}_b}\right],
\end{equation}
where
\begin{equation}
\frac{\mathcal{H}}{\mathcal{H}_b} = 4\,\left[
	\left( \frac{\tau}{\tb} \right)^2 + 
	\left( \frac{\vv }{\vb} \right)^2
\right],
\end{equation}
$\mathcal{H}_b$ is the Hamiltonian of the boundary particle, and, we omit the 
Heaviside step function understanding the domain where
$f_{\text{HP}_0}(\tau,\vv)$ is properly defined.

%------------------------------------------------------------------------------%
This notation assumes a generalized Hofmann-Pedersen distribution of the $n$-th 
order
\begin{equation}
f_{\text{HP}_{n}}(\tau,\vv) =
	\frac{2\,(2\,n+1)}{\pi\,\tb\vb}
	\left(1-\frac{\mathcal{H}}{\mathcal{H}_b}\right)^{n-1/2},
\end{equation}
for $n=0,\frac{1}{2},1,\frac{3}{2},\ldots$.
Independently of the order, the temperature function is parabolic
\begin{equation}
u_{\text{HP}_{n}}^2(t) =
	\frac{\vb^2}{8(n+1)} \left(  1 - 4\,\frac{\tau^2}{\tb^2}\right),
\end{equation}
while the line density is given by
\begin{equation}
\rho_{\text{HP}_{n}}(t) = \frac{2\,\Gamma[n+3/2]}{\tb\sqrt{\pi}\,\Gamma[n+1]}
	\left( 1 - 4\,\frac{\tau^2}{\tb^2} \right)^n,
\end{equation}
so the conventional Hofmann-Pedersen distribution~\cite{hofmann1979bunches}, 
with its parabolic line density, can be addressed as $f_{\text{HP}_{1}}$.

%------------------------------------------------------------------------------%
In this paper, in addition to the SSCHP$_0$ case which admits the analytical 
expressions of the matrix elements, we also consider more realistic models with
$n=1/2,1$ (referred to as SSCHP$_{1/2,1}$ respectively) for a bunch in a 
parabolic potential well, where integrals of the matrix elements were computed 
numerically;
the SSCHP$_{1/2}$ model sometimes is referred as the
{\it waterbag distribution}.
Table~\ref{tab:SSC} contains exact expressions for longitudinal distribution, 
temperature and density functions for all models under consideration.

%------------------------------------------------------------------------------%
\begin{figure*}[h!]
\includegraphics[width=\linewidth]{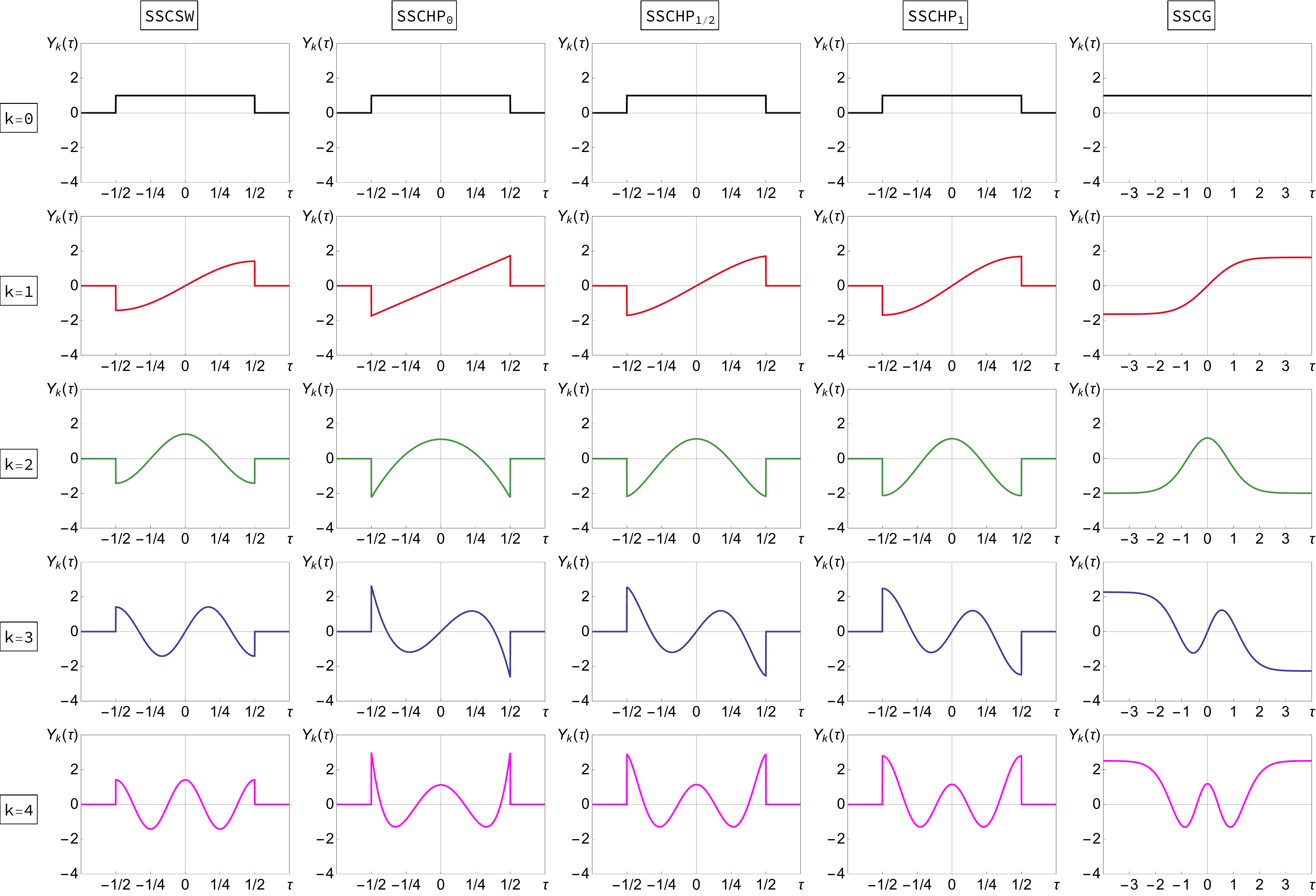}
\caption{\label{fig:SSCHarm}
	First five SSC harmonics $Y_k(\tau)$ for a bunch in a square potential 
	well, bunch in a parabolic potential well (generalized Hofmann-Pedersen
	distributions with $n=0,1/2,1$) and the Gaussian bunch.
	The harmonics are presented in dimensionless units:
	$\tau$ is measured in units of $\tb$ or in units of $\cb$ for the SSCG 
	case.
	}
\end{figure*}
%------------------------------------------------------------------------------%

%==============================================================================%
\begin{table*}[h!]
\caption{\label{tab:SSC}
	Longitudinal distribution function $f(\tau,\vv)$, the average of the
	longitudinal velocity squared $u^2(\tau)$, and, the normalized line 
	density $\rho(\tau)$ for SSC models;
	last row shows graphs of the line density in dimensionless units.
	}
\begin{ruledtabular}
\begin{tabular}{p{1cm}|p{3.2cm}p{3.2cm}p{3.2cm}p{3.2cm}p{3.2cm}}
%------------------------------------------------------------------------------%
								&
SSCSW								&
SSCHP$_{0}$							&
SSCHP$_{1/2}$							&
SSCHP$_{1}$							&
SSCG								\\\hline
	&	&	&	&	&			\\[-0.25cm]
%------------------------------------------------------------------------------%
$f(\tau,\vv)$							&
$\ds\frac{1}{\tb}\text{H}\left[1 - 4\,\frac{\tau^2}{\tb^2}\right]
\mathrm{V}(\vv)$						&
$\ds\frac{2}{\pi\,\tb\vb}
\left(1-\frac{\mathcal{H}}{\mathcal{H}_b}\right)^{-1/2}$	&
$\ds\frac{4}{\pi\,\tb\vb}\text{H}
\left[1-\frac{\mathcal{H}}{\mathcal{H}_b}\right]$		&
$\ds\frac{6}{\pi\,\tb\vb}
\left(1-\frac{\mathcal{H}}{\mathcal{H}_b}\right)^{1/2}$		&
$\ds\frac{e^{-(\tau^2/\cb^2+\vv^2/\vb^2)/2}}
{2\,\pi\,\cb\vb}$						\\[0.5cm]
%------------------------------------------------------------------------------%
$u^2(\tau)$							&
$\ds\vb^2 \equiv \int_\text{SB} \vv^2\,\mathrm{V}(\vv)\,\dd \vv$&
$\ds\frac{\vb^2}{8}\,\left( 1-4\,\frac{\tau^2}{\tb^2} \right)$	&
$\ds\frac{\vb^2}{12}\,\left( 1-4\,\frac{\tau^2}{\tb^2} \right)$	&
$\ds\frac{\vb^2}{16}\,\left( 1-4\,\frac{\tau^2}{\tb^2} \right)$	&
$\ds\vb^2$							\\[0.25cm]
%------------------------------------------------------------------------------%
$\rho(\tau)$							&
$\ds\frac{1}{\tb}$						&
$\ds\frac{1}{\tb}$						&
$\ds\frac{4}{\pi\,\tb}\sqrt{1-4\,\frac{\tau^2}{\tb^2}}$		&
$\ds\frac{3}{2\,\tb}\left( 1-4\,\frac{\tau^2}{\tb^2} \right)$	&
$\ds\frac{e^{-\tau^2/(2\,\cb^2)}}{\sqrt{2\,\pi}\,\cb}$	\\[0.5cm]
%------------------------------------------------------------------------------%
								&
\centering\includegraphics[width=0.36\columnwidth]{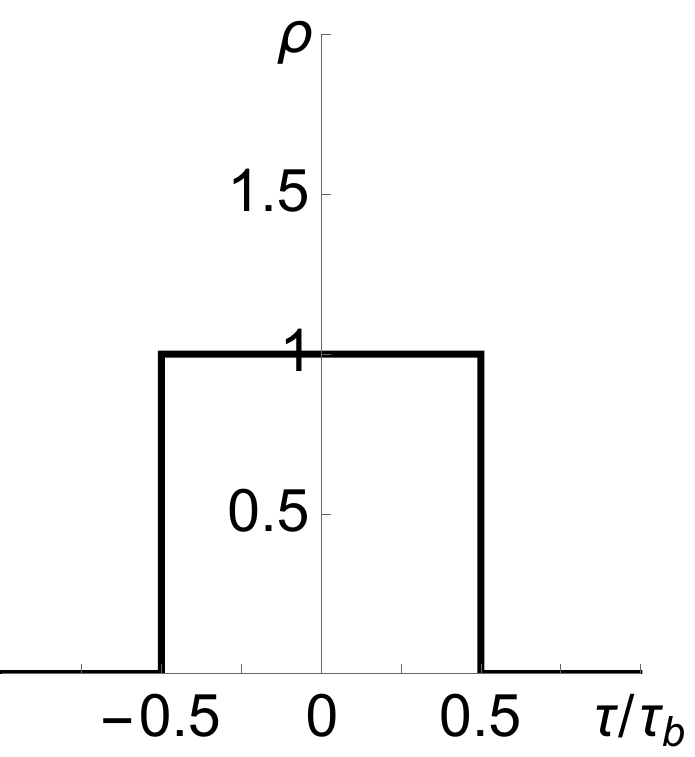}	&
\centering\includegraphics[width=0.36\columnwidth]{small1.pdf}	&
\centering\includegraphics[width=0.36\columnwidth]{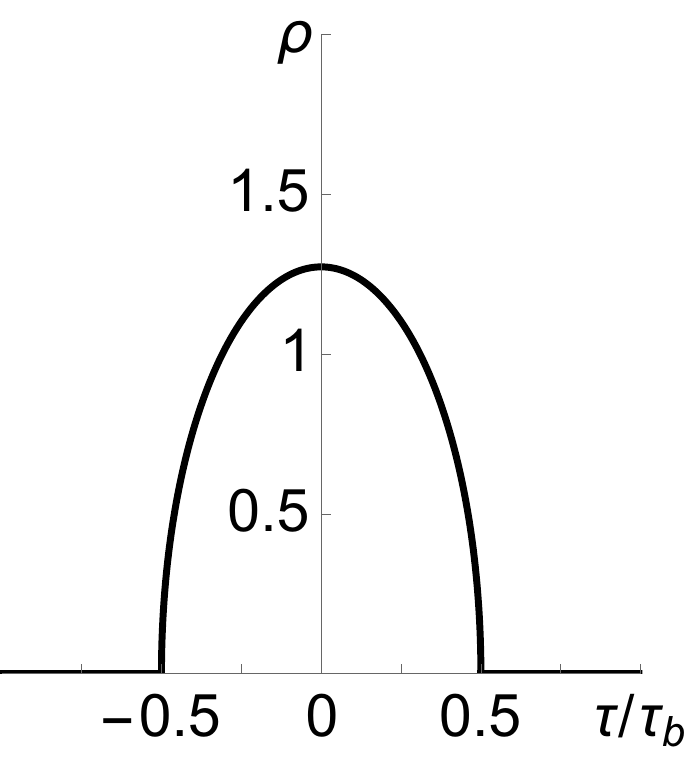}	&
\centering\includegraphics[width=0.36\columnwidth]{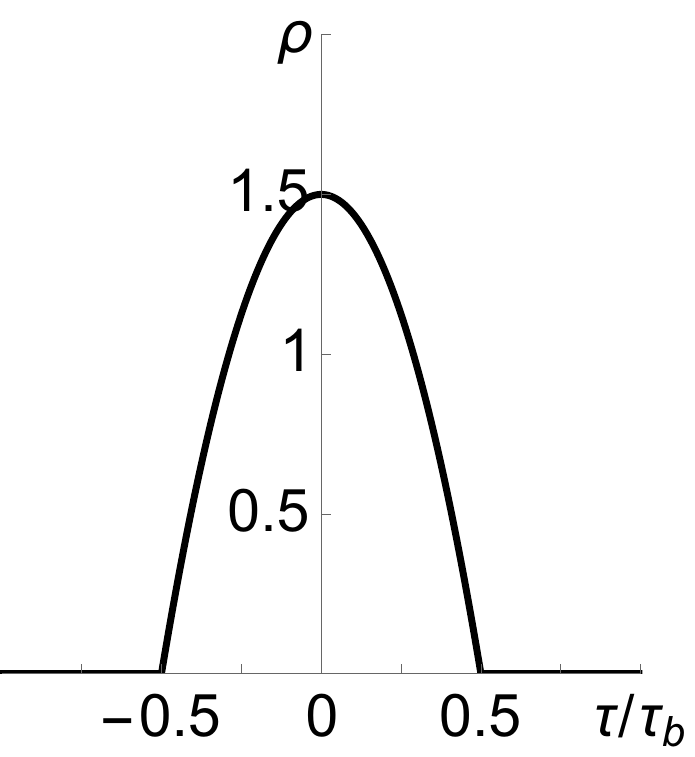}	&
\centering\includegraphics[width=0.36\columnwidth]{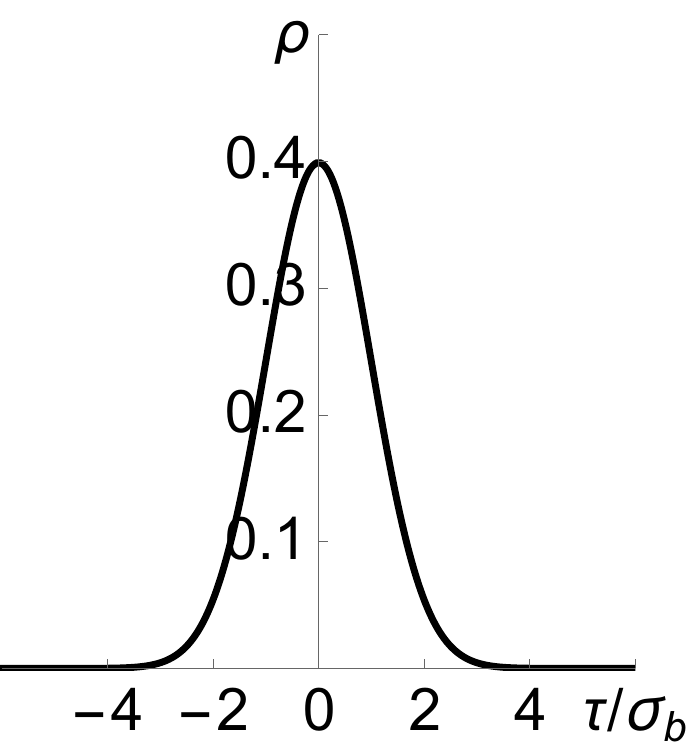}
%------------------------------------------------------------------------------%
\end{tabular}
\end{ruledtabular}
\end{table*}
%==============================================================================%

%------------------------------------------------------------------------------%
The equation for SSC harmonics can be conveniently written in the normalized 
variables, when $\tau$ is measured in units of $\tb$ and $\nu$ in units of 
$\vb^2/[\tb^2\Qef(0)]$, 
\begin{equation}
\label{math:HPHarm}
\left\{
\begin{array}{l}
	\ds Y ''-
	\frac{8\,\tau\,Y'}{1-4\,\tau^2}  +
	\frac{8\,\nu\,(n+1)\,Y}{(1-4\,\tau^2)^{1-n}}  = 0,	\\[0.5cm]
	\ds Y'(\pm 1/2) = 0.
\end{array}
\right.
\end{equation}
Note that $\nu$ is measured in different units in SSCSW and SSCHP;
this choice was made in order to have first eigenvalues $\nu_1 \approx 1$ for 
all the models (for SSCSW and SSCHP$_0$ $\nu_1 = 1$) which simplifies the 
comparison.

%------------------------------------------------------------------------------%
When $n=0$, the S-L problem~(\ref{math:HPHarm}) doesn't have solutions, due to 
the singularities at $\tau = \pm 0.5$.
One way to overcome this problem is to add a small regularization term 
$\epsilon>0$,
\begin{equation}
	Y''(\tau)
	- \frac{8\,\tau}{1+\epsilon-4\,\tau^2}\,Y'(\tau)
	+ \frac{8\,\nu }{1+\epsilon-4\,\tau^2}\,Y (\tau)
	= 0,
\end{equation}
thus removing the singularity within the same bunch length $[-0.5;0.5]$.
With this regularization, the S-L solutions do exist, and, as
$\epsilon \rightarrow 0$, the eigenvalues converge to the
{\it triangular numbers}
\begin{equation}
	\nu_k^{\text{HP}_0} = \left\{k\,(k+1)/2\,|\,k\in\mathbb{Z} \right\} =
	\left\{ 0,1,3,6,10,\ldots \right\}.
\end{equation}
The eigenfunctions converge to functions proportional to the Legendre 
polynomials $P_k(\tau)$, 
\begin{equation}
\label{math:YkHP0}
Y_k^{\text{HP}_0} (\tau)= 
	(-1)^{\left\lfloor{k/2}\right\rfloor}\sqrt{2\,k+1}\,P_k(2\,\tau)
\end{equation}
where the floor function $\left\lfloor n \right\rfloor$ is the integer part of 
$n$.
The $(-1)^{\left\lfloor k/2 \right\rfloor}$ factor guarantees that each even 
(odd) harmonic has cosine-like (sine-like) properties at the origin, see 
Eq.~(\ref{math:SC-like}).

%------------------------------------------------------------------------------%
Another way to find eigenvalues is to look for an even and odd functions with
\begin{equation}
\label{math:SC-like}
	\left\{
	\begin{array}{l}
		Y\phantom{'}(0)	> 0,		\\
		Y'(0)		= 0,
	\end{array}
	\right.
	\qquad\text{or}\qquad
	\left\{
	\begin{array}{l}
		Y\phantom{'}(0)	= 0,		\\
		Y'(0)		> 0,
	\end{array}
	\right.
\end{equation}
respectively.
When the small regularization term $\epsilon > 0$, the derivatives $Y'(\pm 0.5)$
vanish for the same triangular values, $\nu_k^{\text{HP}_0}$.
With $\epsilon = 0$, only the triangular eigenvalues $\nu = k(k+1)/2$ yield the 
solutions, remaining finite at the bunch edges $\tau = \pm 0.5$.

%------------------------------------------------------------------------------%
The first ten numerically obtained values of $\nu_k$ for SSCHP$_{1/2,1}$ are
given in Table~\ref{tab:SSCnu} and first five harmonics for all cases are 
presented in Fig.~\ref{fig:SSCHarm}.

%==============================================================================%
\begin{table}[th!]
\caption{\label{tab:SSCnu}
	First ten eigenvalues $\nu_k$ for SSC models: bunch inside a square
	(SSCSW) and parabolic (SSCHP$_{0,1/2,1}$) potential wells,
	and, Gaussian bunch (SSCG).
	}
\begin{ruledtabular}
\begin{tabular}{lccccc}
$k$	& SSCSW & SSCHP$_0$ & SSCHP$_{1/2}$ & SSCHP$_1$ & SSCG	\\\hline
	&	&	&	&	&	\\[-0.25cm]
0	& 0	& 0	& 0	& 0	& 0	\\
1	& 1	& 1	& 1.1002& 1.1555& 1.342	\\
2	& 4	& 3	& 3.378	& 3.5910& 4.3245\\
3	& 9	& 6	& 6.8078& 7.2713& 8.8978\\
4	& 16	& 10	&11.386	&12.1905&15.0531\\
5	& 25	& 15	&17.1115&18.3465&22.7868\\
6	& 36	& 21	&23.9837&25.7383&32.0966\\
7	& 49	& 28	&32.0023&34.3653&42.9817\\
8	& 64	& 36	&41.1672&44.2272&55.441	\\
9	& 81	& 45	&51.4783&55.3235&69.474	
\end{tabular}
\end{ruledtabular}
\end{table}
%==============================================================================%

%==============================================================================%
%==============================================================================%
\subsubsection{\label{sec:SSCG}Gaussian bunch}

%------------------------------------------------------------------------------%
The last strong space charge case we present is the model of a thermalized beam 
with Gaussian distribution function (SSCG):
\begin{equation}
	f_\text{G}(\tau,\vv) =
		\frac{1}{2\,\pi\,\cb\vb}
		\exp\left[
			-\frac{\tau^2}{2\,\cb} - \frac{\vv^2}{2\,\vb}
		\right].
\end{equation}
The normalized line density is simply a Gaussian distribution
\[
	\rho_\text{G}(\tau) = \frac{1}{\sqrt{2\,\pi}\cb}
	\exp \left[
		-\frac{\tau^2}{2\,\cb}
	\right],
\]
and as in the case of SSCSW the average square of velocity is constant along 
the bunch
\begin{equation}
	u_\text{G}^2(\tau) = \vb^2.
\end{equation}
Using the similar normalized variables as for the SSCHP cases, where $\tau$ is 
measured in units of $\cb$ and $\nu$ in units of
$Q_s^2/\Qef(0) \equiv \vb^2/[\cb^2\Qef(0)]$, the eigenfunction equation for 
transverse bunch oscillations is
\begin{equation}
\label{math:SSCG}
\left\{
	\begin{array}{l}\displaystyle
		Y''(\tau) + \nu\,e^{-\tau^2/2} \,Y(\tau) = 0,	\\[0.25cm]
		Y'(\pm \infty) = 0.
	\end{array}
\right.
\end{equation}
Below, we will adopt the convention that for all practical purposes the bunch
length is defined as $\tb = 3\,\cb$.
The first 10 numerically obtained eigenvalues $\nu_k$ are listed in the 
Table~\ref{tab:SSCnu} and plots of first few eigenfunctions are given in 
Fig.~\ref{fig:SSCHarm}.

%==============================================================================%
%------------------------------------------------------------------------------%
%==============================================================================%
\subsection{\label{sec:AB}Airbag in a square well}

%------------------------------------------------------------------------------%
In addition to SSC cases, we consider the airbag longitudinal distribution
inside a square well, abbreviated as ABS (AirBag Square well) model,
\begin{equation}
	f_\mathrm{ABS}(\tau,\vv) \propto
	\left[ \delta(\vv-\vv_0) + \delta(\vv+\vv_0) \right]
	\text{H}\left[ 1-4\,\frac{\tau^2}{\tb^2}\right],
\end{equation}
suggested by M.~Blaskiewicz \cite{blaskiewicz1998fast}.
The ABS model allows the bunch spectrum to be determined for a wide class of
wake functions in the presence of arbitrary space charge tune shift without 
requiring the expansion in basis functions, thereby avoiding convergence 
difficulties.

%------------------------------------------------------------------------------%
It is convenient to take for this model the following convention
$\tau\in[-\tb/2;\tb/2]$ for the position along the bunch, where all particles 
move with the same absolute value of the velocity,  
$\dd\,\tau/\dd\theta = \pm \vv_0$.
Transverse offsets in the two fluxes of particles $X_\pm(\theta,\tau)$ can be 
looked for as
\begin{equation}
	X_\pm(\theta,\tau) = e^{-i\,(\Qb+\Qx)\theta} x_\pm(\tau)
\end{equation}
where $\Qb$ is the bare betatron tune and $\Qx$ is the tune shift to be found. 
Then, equations for the amplitudes along the bunch $x_\pm(\tau)$ are given by
\begin{eqnarray}
\label{math:xp}
	\frac{\dd\,x_+}{\dd \tau} & = & \frac{i}{\vv_0} \left[
		\left( \frac{\Qsc}{2}+\Qx \right) x_+ - \frac{\Qsc}{2}\,x_- - F
	\right],\qquad						\\
\label{math:xm}
	\frac{\dd\,x_-}{\dd \tau} & = & \frac{i}{\vv_0} \left[
		\frac{\Qsc}{2}\,x_+ - \left( \frac{\Qsc}{2}+\Qx \right) x_- + F
	\right],\qquad
\end{eqnarray}
with the boundary conditions
\begin{equation}
\begin{array}{c}
	x_+( \tb/2) = x_-( \tb/2),	\\[0.2cm]
	x_+(-\tb/2) = x_-(-\tb/2).
\end{array}
\end{equation}

%------------------------------------------------------------------------------%
The force is defined by a wake function
\begin{equation}
	F(\tau) = \kappa\,\int_\tau^{\tb/2}
		W(\tau-\sigma)\,\overline{x}(\sigma)
	\,\dd\sigma
\end{equation}
and satisfies an integro-differential equation:
\begin{equation}
	\label{math:F(t)}
	\frac{\dd\,F(\tau)}{\dd \tau} =-\kappa\,W(0)\,\overline{x}(\tau) +
		\kappa\,\int_\tau^{\tb/2} \frac{\pd}{\pd \tau} W(\tau-\sigma) 
	\overline{x}(\sigma)
		\,\dd \sigma
\end{equation}
with $\overline{x} = (x_+ + x_-)/2$.
For a wake function in the form
\begin{equation}
\label{math:W(tau)}
	W(\tau) = -W_0\,\sum_{k=1}^{n} C_k\,e^{\alpha_k\,\tau}
\end{equation}
the force is
\begin{equation}
	F = \sum_{k=1}^n F_k =-\kappa\,W_0  \sum_{k=1}^n  C_k \int_\tau^{\tb/2} 
	e^{\alpha_k(\tau-\sigma)}\,\overline{x}(\sigma)\,\dd \sigma,
\end{equation}
and for every $k=1,\ldots,n$,
\begin{equation}
\label{math:dFk}
	\frac{\dd\,F_k(\tau)}{\dd \tau} =
	\frac{\kappa\,W_0}{2}\,C_k(x_+ + x_-) + \alpha_k\,F_k,
\end{equation}
with $F_k(\tb/2) = 0$.

%------------------------------------------------------------------------------%
Measuring $\tau$ in units of $\tb$ and defining
\begin{equation}
	f(\tau) = F(\tau)/\Qs
\end{equation}
where $\Qs = \pi\,\vv_0/\tb$ is the synchrotron tune,
Eqs.~(\ref{math:xp},\ref{math:xm}) and (\ref{math:dFk}) can be presented as a 
set of ordinary linear homogeneous differential equations,
\begin{equation}
	\frac{\dd\,\mathrm{U}}{\dd\tau} = \mathrm{M}\,\mathrm{U},
	\label{dUdtau}
\end{equation}
with boundary conditions
\begin{eqnarray}
\label{math:bc1}
	f_k( 1/2) &=& 0,			\\
\label{math:bc2}
	x_+( 1/2) &=& x_-( 1/2),		\\
\label{math:bc3}
	x_+(-1/2) &=& x_-(-1/2),
\end{eqnarray}
where the vector $\mathrm{U} = (x_+,x_-,f_1,f_2,\ldots)$ has $n+2$ components 
and the matrix $\mathrm{M}$ is combined from the coefficients of
Eqs.~(\ref{math:xp},\ref{math:xm},\ref{math:dFk}).
To solve this set of equations, we suggest an algorithm different from one 
applied by M.~Blaskiewicz.

%------------------------------------------------------------------------------%
The differential Eq.~(\ref{dUdtau}), first, can be transformed into a 
difference one,
\begin{equation}
	\mathrm{U}_{n+1} - \mathrm{U}_n = \Delta\tau\,\mathrm{M}\,\mathrm{U}_n.
\end{equation}
Choosing the number of steps to be $n=2^m$ and $\Delta\tau = 1/2^m$, one has
\begin{equation}
	\mathrm{U}_0 \equiv \mathrm{U}(-1/2) = \left(
		\frac{1}{2^m}\,\mathrm{M} + \mathrm{I} 
	\right)^{-2^m}\,\mathrm{U}_n,
\end{equation}
where $\mathrm{U}_0$ and $\mathrm{U}_n$ are values of $\mathrm{U}$ at the tail 
and the head of the bunch respectively, and, according to the boundary 
conditions (\ref{math:bc1}--\ref{math:bc3})
\[
	\mathrm{U}_n \equiv \mathrm{U}(1/2) = (1,1,0,0,\ldots,0).
\]
Those values of $\Qx$, with which the boundary condition at the tail of the 
bunch is satisfied, i.e. 
\begin{equation}
	x_+^{(0)} - x_-^{(0)} = 0,
\end{equation}
constitute the bunch spectrum;
they can be found by a proper scan of the complex plane of $\Qx$.
However, for the instability thresholds, it is sufficient to scan only real
$\Qx$: the threshold can be detected as reduction of the number of these real 
modes by two.

%------------------------------------------------------------------------------%
This algorithm proves to be extremely time-efficient; the solution always
converges with the number of intervals $1/\Delta \tau$, while the CPU time 
grows only logarithmically, $\propto \log_2 (1/\Delta \tau) =m$.

\begin{figure}[h!]
\includegraphics[width=\linewidth]{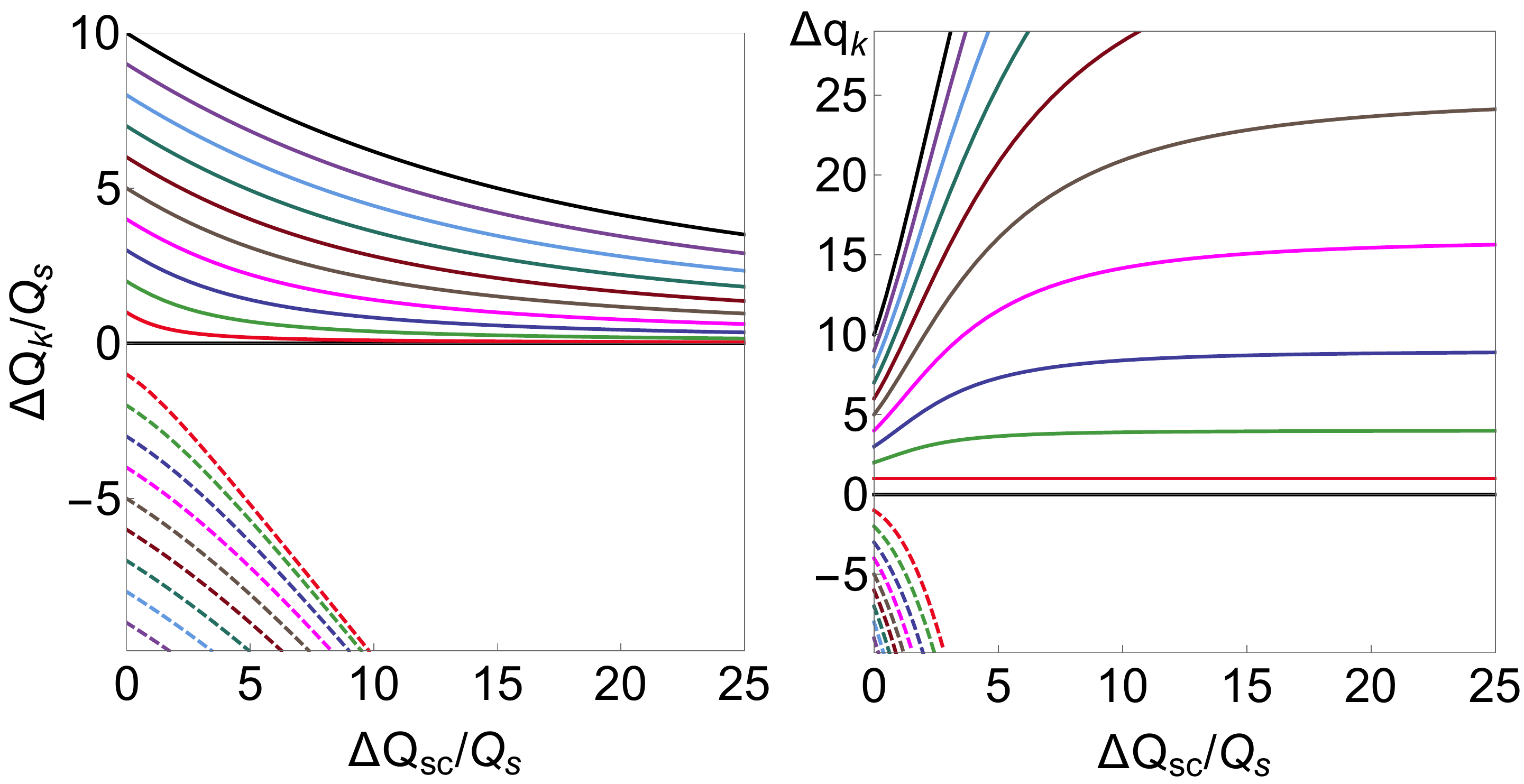}
\caption{\label{fig:ABNoWake}
	Eigenvalues (left) and normalized eigenvalues (right) for the ABS model 
	as functions of the SC parameter for no-wake case, $k=-9,\ldots,10$.
	Positive and negative parts of spectrum are shown with solid and
	dashed lines respectively.
	}
\end{figure}

%------------------------------------------------------------------------------%
For the no-wake case, the ABS spectrum is calculated as
\begin{equation}
\label{math:ABSNoWake}
	\frac{\Qx}{\Qs} = -\frac{\Qsc}{2\,\Qs} \pm \
	\sqrt{\left(\frac{\Qsc}{2\,\Qs}\right)^2 + k^2},
\end{equation}
for $k = 1, 2,3,\ldots$ and additional zeroth mode
\[
	\Delta Q_0 = 0
\]
which is not affected by the SC.
Without the SC, it results in the equidistant spectrum, $\Delta Q_k = k\,\Qs$.
When the SC becomes strong,
\begin{equation}
\label{math:SSC}
	\Qsc \gg 2\,k\,\Qs 
\end{equation}
the spectrum separates on the positive part, quadratic with the mode number:
\[
	\Delta Q_k = k^2 \Qs^2 / \Qsc,
\]
and the negative part, with
\[
	\Delta Q_k \approx -\Qsc - k^2 \Qs^2/\Qsc,
\]
see Fig.~\ref{fig:ABNoWake}.
When the space charge parameter gets larger and larger, all the modes of each 
group become degenerate, with zero tune shift for the positive modes and $-\Qsc$
for the negative ones.
This degeneracy can be understood in the following way.
Without wake, the ABS equations of motion do not change after the 
transformation 
$\tau \rightarrow \tb - \tau$, $x_+ \leftrightarrow x_-$.
If the SC is so strong, that the synchrotron motion can be completely 
neglected, there is an additional symmetry between $\tau$ and $\tb - \tau$. 
Thus, at that extremely high SC, the modes must be either stream-even, 
$x_+(\tau) = x_-(\tau)$, or stream-odd,  $x_+(\tau) = -x_-(\tau)$.
Due to the boundary conditions, stream-odd modes are zeroed at the edges, while 
stream-even ones have zero derivatives there.
For the stream-even modes, when the two streams oscillate in phase, the tune 
shift is obviously zero, while for the stream-odd modes all the tunes have to 
be 
identically shifted down by the space charge tune shift.

In order to make a comparison with the SSC case, we use normalized tunes
\begin{equation}
	\qx = \Qx / \Delta Q_1,
\end{equation}
so that the distance between the first and zeroth modes is equal to one for any 
value of the SC parameter, $\Qef(0)/\Qs$, when there is no wake.
When condition~(\ref{math:SSC}) is satisfied, ABS positive spectrum coincides 
with the SSCSW one:
\begin{equation}
	\Qx = k^2 \Qs^2/\Qsc , \qquad k=0,1,2,\ldots. ;
\end{equation}
for the SSCSW model, all the details of the longitudinal phase space density 
may play a role only by means of the average synchrotron frequency $\Qs$.

%------------------------------------------------------------------------------%
%------------------------------------------------------------------------------%

%\newpage
%==============================================================================%
%------------------------------------------------------------------------------%
%------------------------------------------------------------------------------%
%------------------------------------------------------------------------------%
%==============================================================================%
\section{\label{sec:NegWakes}Negative wakes}

%------------------------------------------------------------------------------%
Without wakes, all ABS modes are divided into two groups, with positive and 
negative tune shifts $\Qx$;
at growing SC, the former tend to zero, while the latter tend to the SC tune 
shift.
The addition of wake fields will shift the tunes of the modes.
It is convenient to number the mode using the same number to which that mode 
corresponds without wakes.
This mode number definition is unambiguous provided that the given mode never 
coupled with its neighbors at smaller wake amplitudes, which is sufficient for 
our purposes.
Following that, we distinguish all the modes at any wake between
{\it positive and negative modes}, or
{\it positive or negative parts of the spectrum},
which is unambiguous provided that there is no coupling between the positive 
and negative modes, which is always the case below.
{\it The reader should not be confused by the fact that wakes may shift tunes of
some positive modes to negative values; we still refer to such modes as
positive.}              

%------------------------------------------------------------------------------%
In this section, constant-sign wakes are considered for the ABS, as well as for 
the SSC problems;
as it follows from the Maxwell equations, this sign can only be negative
\cite{chao1993physics}.
We limit ourselves here by delta-functional, constant, step-function,
exponential and resistive wall wakes.
All the wakes are assumed to be causal.
Sometimes, the tunes at a given wake and SC are convenient to present in units 
of the first eigenvalue at the same SC and zero wake.
To facilitate this, we define 
\begin{equation}
	\kappa^* = \kappa \Big/ \left[-\frac{\Qsc}{2\,\Qs} + \
	\sqrt{\left(\frac{\Qsc}{2\,\Qs}\right)^2 + 1}\right].
\end{equation}
In these units, the positive part of the spectrum shows scale invariance for 
large values of $\Qsc$, Eq.~(\ref{math:SSC}).
In order to describe the instability, we introduce a dimensionless {\it wake
parameter} $\chi$ and its normalized value $\chi^*$:
\begin{equation}
\label{math:chi2}
\chi   = \frac{\kappa  \,W_0}{\Qs}
\qquad\text{and}\qquad
\chi^* = \frac{\kappa^*\,W_0}{\Qs},
\end{equation}
where $\tb$ and $W_0$ are the bunch length and the wake amplitude specified for 
every case separately ($\tb = 3\cb$ for SSCG).
When the SC is getting strong, Eq.~(\ref{math:chi2}) corresponds to
Eq.~(\ref{math:chi}) with $\Qef(0)/\Qs^2$ being understood as units of coherent
tune shifts.

%==============================================================================%
%------------------------------------------------------------------------------%
%==============================================================================%
\subsection{\label{sec:DeltaWake}Delta wake}

\begin{figure*}[h!]
\includegraphics[width=0.9\linewidth]{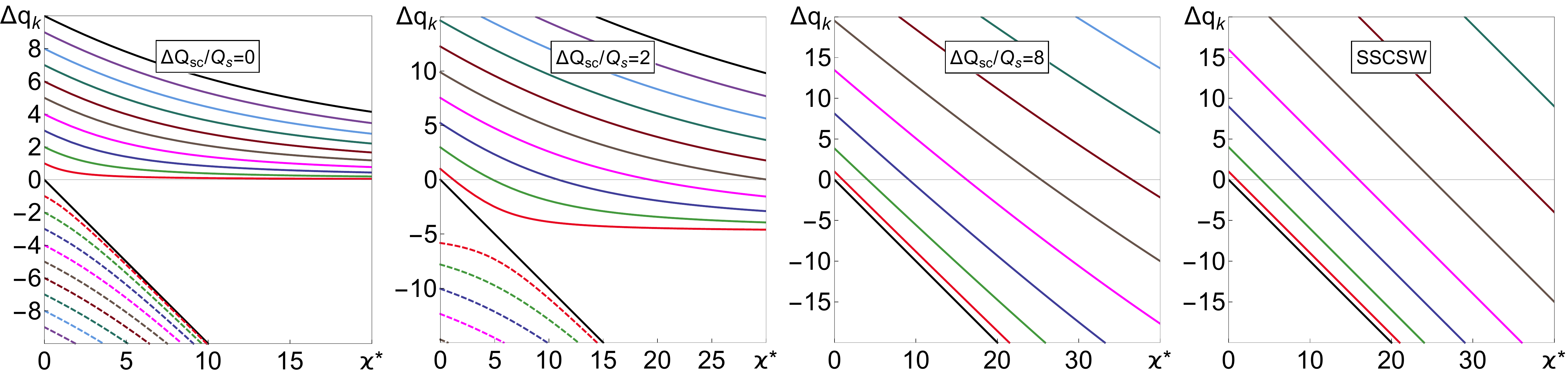}
\includegraphics[width=0.89\linewidth]{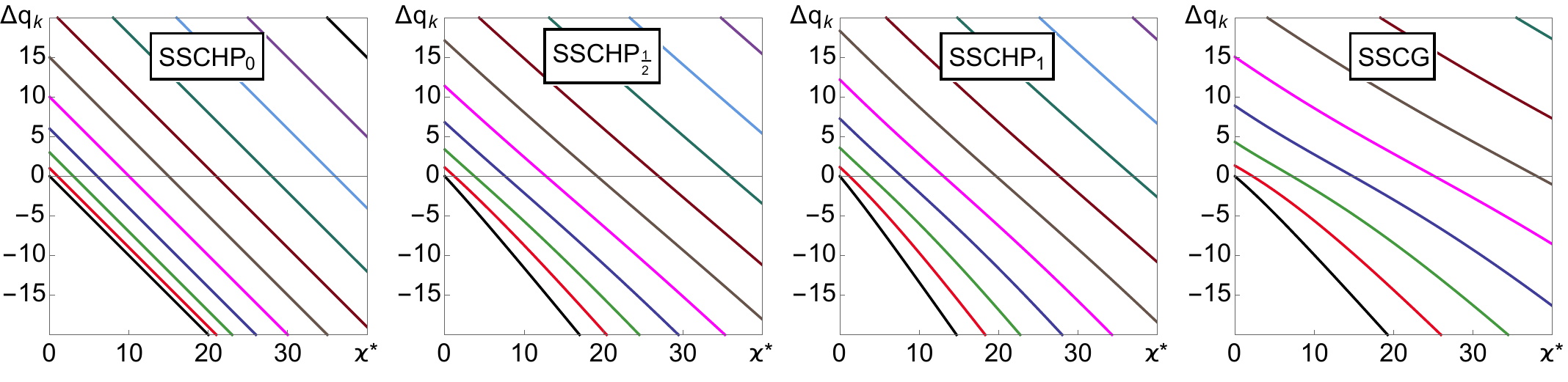}
\caption{\label{fig:DeltaWake}
	Top row: normalized tune shifts for the ABS model and delta wake for 
	different values of the SC parameter;
	the last plot in a row is the spectrum at the SSC limit --- SSCSW.
	Bottom row: normalized spectra for remaining SSC models.
	}
\end{figure*}

%------------------------------------------------------------------------------%
Our first negative-wake example is one of image charges, the delta wake,
\begin{equation}
	W(\tau) = - W_0\tb\,\delta(\tau).
\end{equation}
In this case the airbag model allows analytical 
solution~\cite{PhysRevSTAB.12.114201}
\begin{equation}
\label{math:ABSDeltaWake}
	\frac{\Qx}{\Qs} = -\frac{\Qsc+\kappa\,W_0}{2\,\Qs} \pm \
	\sqrt{\left(\frac{\Qsc-\kappa\,W_0}{2\,\Qs}\right)^2 + k^2},
\end{equation}
for $k = \pm 1,\pm 2,\ldots$ and the zeroth mode
\[
	\Delta Q_0/\Qs =-\kappa\,W_0.
\]
The system is stable for any values of wake amplitude and SC.
The first row of Fig.~\ref{fig:DeltaWake} shows normalized tune shift for ABS 
model as a function of the wake parameter for different values of SC.
%------------------------------------------------------------------------------%
As the SC parameter is increased, the positive and the negative parts of the 
spectrum begin to separate, and, at very high SC parameter the bunch spectra 
corresponds to the one of SSCSW
(last plot in the first row of Fig.~\ref{fig:DeltaWake}).

%------------------------------------------------------------------------------%
%------------------------------------------------------------------------------%

%------------------------------------------------------------------------------%
\begin{figure*}[h!]
\includegraphics[width=0.71\linewidth]{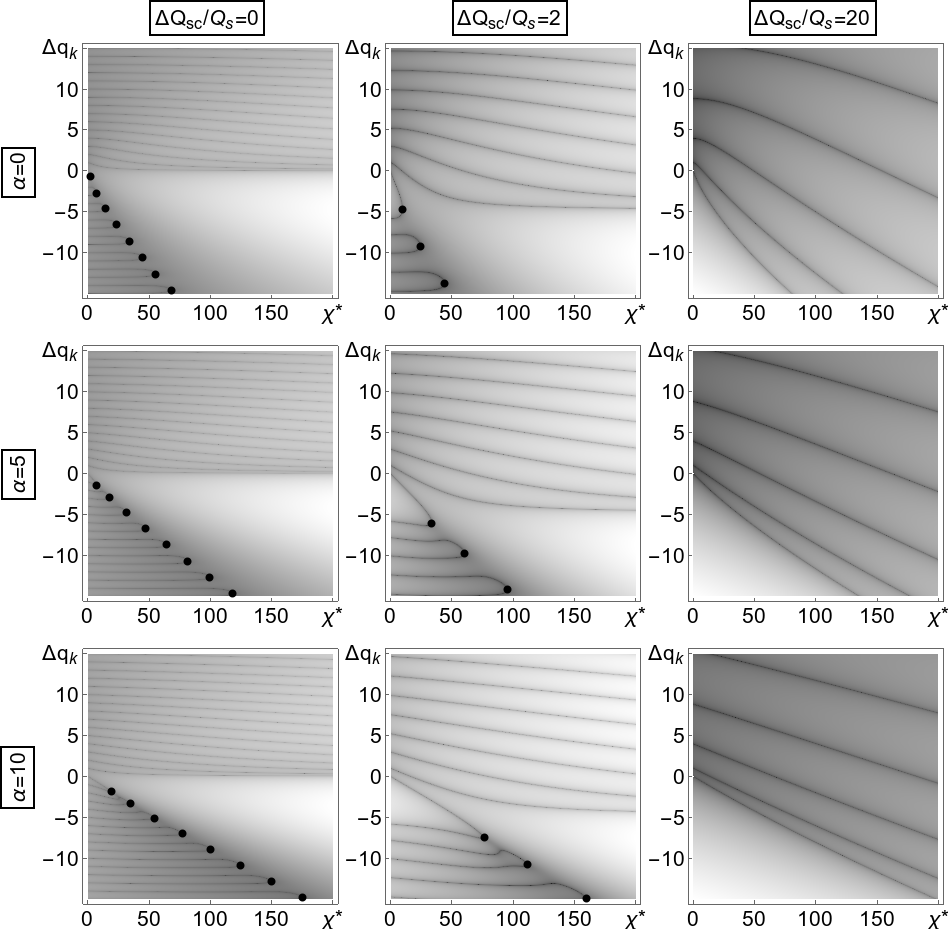}
\includegraphics[width=0.228\linewidth]{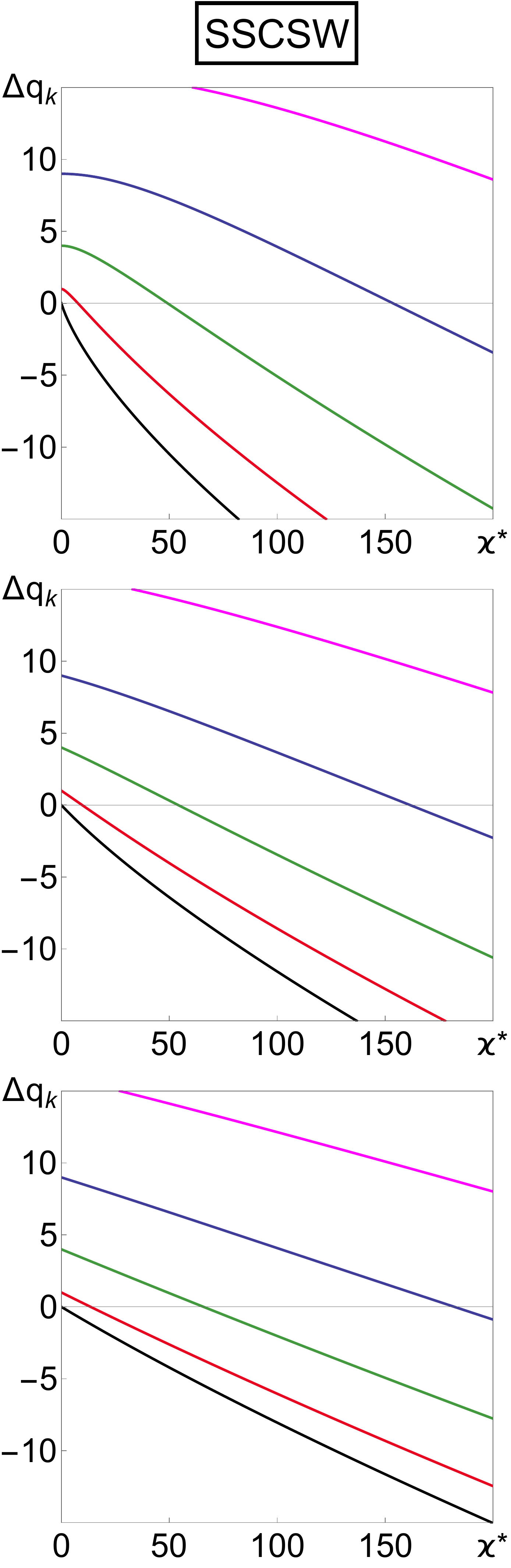}
\caption{\label{fig:ABExpWakeSpectra}
	Spectra for the ABS model and constant ($\alpha = 0$) or exponential
	($\alpha = 5,10$) wakes.
	Different columns correspond to different values of the space 
	charge parameter $\Qsc/\Qs=0,2,20$.
	The TMCI thresholds are shown by black points.
	The last plot in each row is the spectrum at the SSC limit --- SSCSW 
	model.
	}
\end{figure*}
%------------------------------------------------------------------------------%

%------------------------------------------------------------------------------%
For the SSC models, matrix elements are given by 
\begin{equation}
	\mW =
	-W_0\,\int_\text{SB} \rho^2(\tau)\,Y_l(\tau)\,Y_m(\tau)\,\dd\tau,
\end{equation}
and were found numerically for the SSCHP$_{1/2,1}$ and SSCG cases.
In the case of  SSCSW and SSCHP$_{0}$ (and any other distribution with constant 
line density), this integral reduces to 
\begin{equation}
	\mW =	-W_0\,\delta_{lm},
\end{equation}
yielding the eigenvalues
\begin{equation}
	\qx = \nu_k + \kappa\,\widehat{\text{W}}_{kk} = \nu_k - \kappa\,W_0.
\end{equation}
The second row of  Fig.~\ref{fig:DeltaWake} shows the normalized tune shift as a
function of wake amplitude for all SSCHP and SSCG models.
The spectra for the more realistic distributions (SSCHP$_{1/2,1}$ and SSCG) are 
similar to that of SSCHP$_{0}$ and SSCSW with the only difference of faster 
deflection of the zeroth mode compared to the other modes.

%------------------------------------------------------------------------------%
The Hamiltonian nature of the system ensures beam stability for delta wakes: 
the delta wake can be taken into account with a term proportional to a double 
sum
\[
	\Sigma_{lm} x_l x_m \delta (\tau_l - \tau_m)
\]
in the total Hamiltonian, 
where $x_l$ and $x_m$ are offsets of individual particles and $\tau_l$, 
$\tau_m$ are their positions inside the bunch.

%==============================================================================%
%------------------------------------------------------------------------------%
%==============================================================================%
\subsection{\label{sec:ExpWake}Exponential and constant wakes}

%------------------------------------------------------------------------------%
As the next example, consider exponential wakes
\begin{equation}
	W(\tau) = -W_0\,e^{\alpha\,\tau},
\end{equation}
including a constant wake, $\alpha=0$.
With these wakes, TMCI problem for the ABS model was formulated and solved by 
M.~Blaskiewicz and for HP$_0$ and SW cases by V.~Balbekov, within the SSC 
approximation
\cite{
balbekov2016dependence,
balbekov2016tmci,
balbekov2017transverse}.
Here we summarize some of these results and make a comparison with more 
realistic model of a Gaussian bunch and SSCHP$_{1/2,1}$.

%==============================================================================%
%==============================================================================%
\subsubsection{ABS model}
%==============================================================================%

\begin{figure}[h!]
\includegraphics[width=0.75\linewidth]{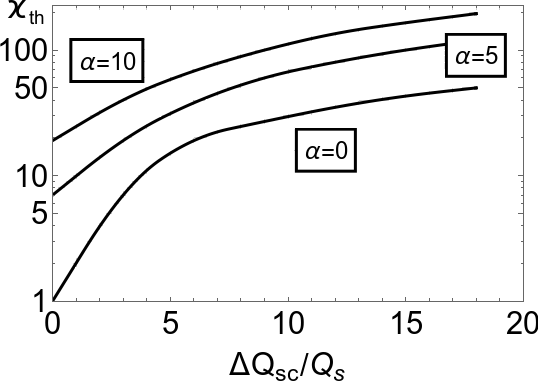}
\caption{\label{fig:ABExpWakeSummary}
	TMCI threshold as a function of SC for the ABS model with the 
	exponential wakes (reproduction of Ref.~\cite{blaskiewicz1998fast}).
	}
\end{figure}

%------------------------------------------------------------------------------%
Figure~\ref{fig:ABExpWakeSpectra} shows the normalized coherent tune shifts for
constant and exponential ($\alpha= 5,10$) wakes for the ABS model.
As one can see, TMCI may occur only at the negative part of the spectrum and the
thresholds monotonically increase with $\Qsc$.
The last two plots in each row show the spectra for large value of SC
($\Qsc/\Qs = 20$) and the one produced from the SSCSW model respectively.
These plots look identical to each other and hence show the scale invariance at 
the SSC limit.
Behavior of the TMCI thresholds for these cases is summarized in 
Fig.~\ref{fig:ABExpWakeSummary} which we reproduce, albeit by another method, 
after M.~Blaskiewicz \cite{blaskiewicz1998fast}.
When the instability threshold increases with SC without limit, the TMCI is 
referred in this paper as {\it vanishing}.

%------------------------------------------------------------------------------%
%------------------------------------------------------------------------------%

%------------------------------------------------------------------------------%
Increase of the TMCI thresholds with space charge was first demonstrated in
Ref.~\cite{blaskiewicz1998fast}, soon after that was explained in 
Ref.~\cite{ng1999stability}:
while the wake deflects down mostly the zeroth mode, and to a lesser extent does
that for the mode $\Delta Q_{-1}$, the space charge does exactly the opposite, 
deflecting down the mode $\Delta Q_{-1}$ and keeping the tune of the mode $0$ 
untouched;
thus, the mode crossing either happens at higher wakes, or does not happen at
all.
However, it was later speculated in Ref.~\cite{burov2009head} that the increase 
of the threshold with the space charge may be non-monotonic, that at 
sufficiently high SC tune shift, the TMCI threshold may start to go down. 
A reason for that speculation was derived from the unlimited reduction of the 
mode separation with the space charge;
when the neighbor modes are closer and closer, it should take less and less to 
couple them.
This speculation was apparently confirmed within the SSC approximation by 
computation of the TMCI threshold versus the SC tune shift with constant and 
resistive-wall wakes, where the non-monotonic behavior of the instability 
threshold was observed.
However, those computations of Ref.~\cite{burov2009head} were made with an 
insufficient number of modes and with limited accuracy of the matrix element 
computation, so the formulated hypothesis remained open.
Some supporting results for that were provided by tracking simulations of
Ref.~\cite{blaskiewicz2012comparing}.
This problem was recently addressed by 
V.~Balbekov~\cite{balbekov2017transverse}.
Specifically, it has been shown by him, that the TMCI threshold computed for 
the exponential wakes increases without limit as more basis functions are taken 
into account.
In other words, for such wakes at SSC case there is no TMCI at all, the
instability may happen only with comparable SC and wake-driven tune shifts, as 
in Fig.~\ref{fig:ABExpWakeSummary}.
Below we confirm that with our results.

%==============================================================================%
%==============================================================================%
\subsubsection{Convergence of SSC models}
%==============================================================================%

\begin{figure}[h!]
\includegraphics[width=\linewidth]{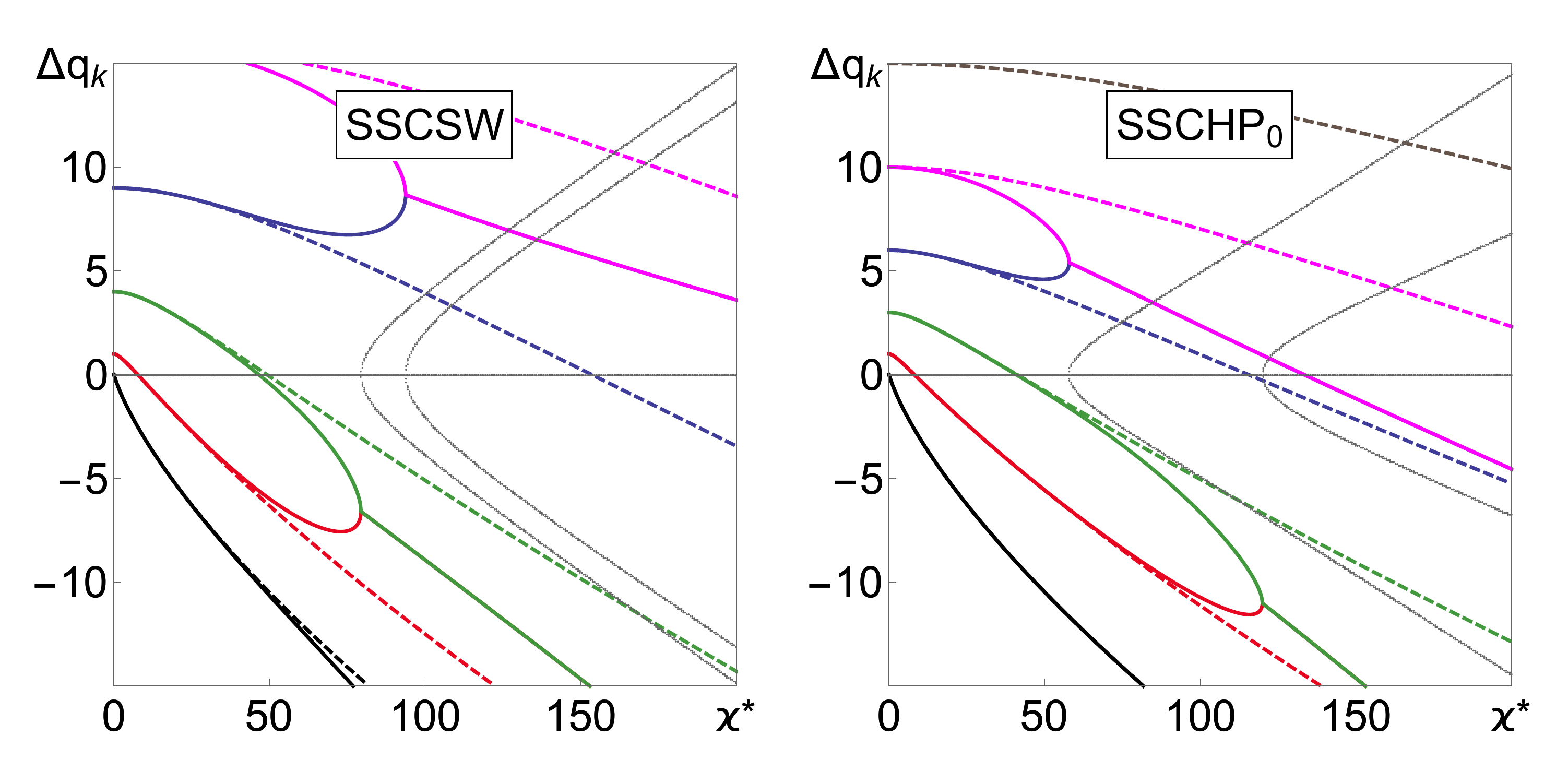}
\caption{\label{fig:SSCConstWakeSpectra}
	Real (shown in color) and imaginary (gray colors) parts of the spectra
	for SSCSW and SSCHP$_0$ with a constant wake.
	Solid and dashed lines are obtained using 5 or 50 modes respectively.
	For sufficiently large number of modes and sufficiently good accuracy of
	the matrix element computations, TMCI vanishes.
	}
\end{figure}

%------------------------------------------------------------------------------%
In this subsection, we will discuss the question of spectra convergence for SSC 
models with respect to the number of modes taken into analysis.
We will use the constant wake model for illustrative purposes.
Figure~\ref{fig:SSCConstWakeSpectra} shows the spectra for SSCSW and SSCHP$_{0}$
for constant wake with different number of modes taken into account.
Solid lines show real and imaginary parts of the spectra with truncation at 5 
harmonics.
Dashed lines are the same spectra computed with 50 basis harmonics.
Although both cases look similar for smaller values of the wake parameter 
$\chi^*$, we clearly see the onset of instability with 5 harmonics, while no 
instability is observed when number of harmonics is sufficiently large.

%------------------------------------------------------------------------------%
%------------------------------------------------------------------------------%

\begin{figure}[h!]
\includegraphics[width=\linewidth]{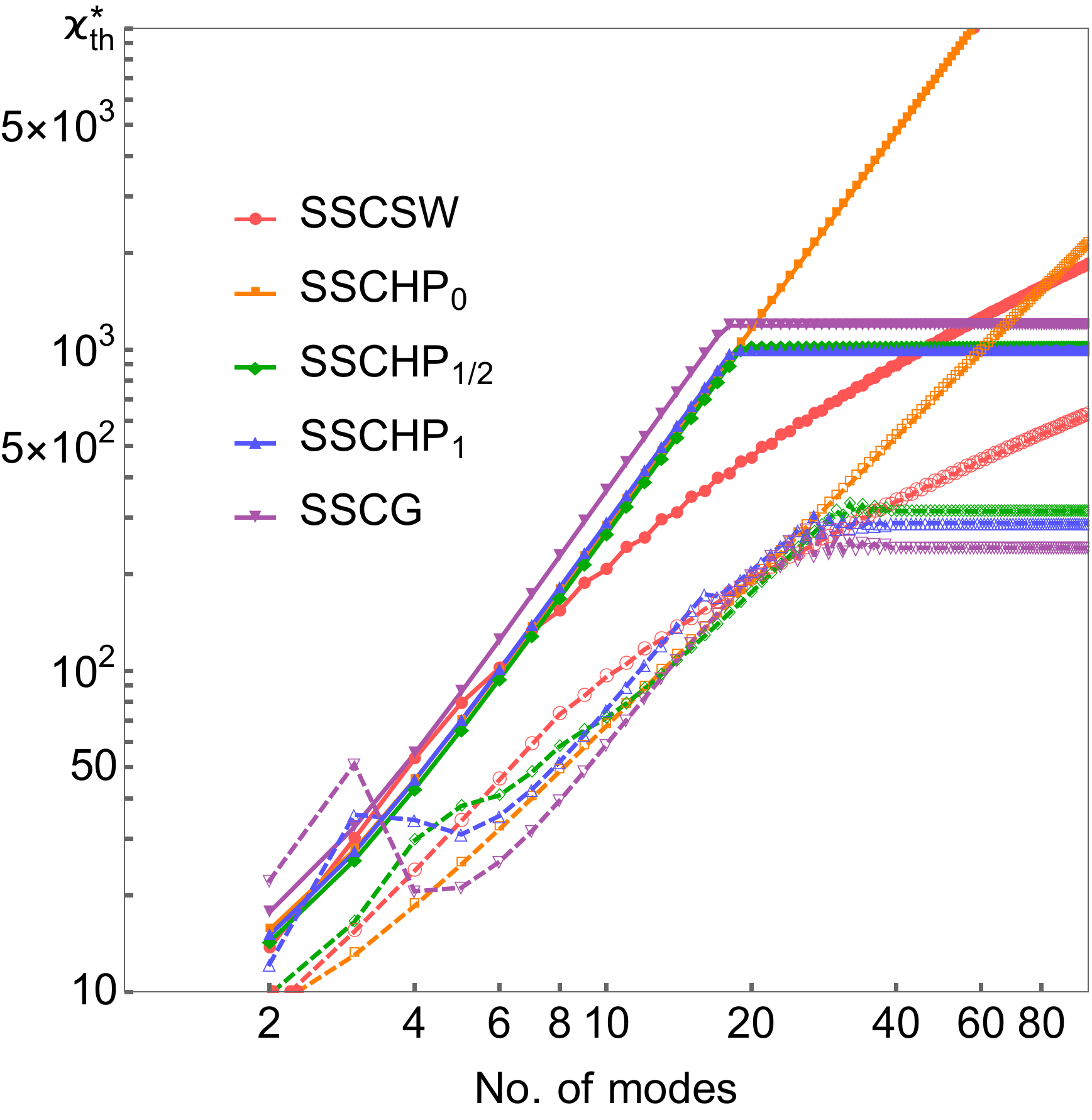}
\caption{\label{fig:SSCNegWakeConv}
	TMCI threshold as a function of number of modes for constant and
	resistive wall wakes (solid and dashed lines respectively) at SSC.
	The unlimited growth with the truncation parameter means that there is
	no TMCI.
	The saturation of the threshold for numerical models (SSCHP$_{1/2,1}$
	and SSCG) does not reflect the instability and is due to limited 
	precision of calculated matrix elements, $\mW$.
	}
\end{figure}

%------------------------------------------------------------------------------%
Figure~\ref{fig:SSCNegWakeConv} summarizes the results stated above.
It demonstrates the growth of TMCI threshold with the truncation parameter, a 
number of harmonics included in the computations.
For all the cases, the thresholds monotonically increase with the mode 
truncation parameter, thus demonstrating that the beam is stable against TMCI. 
The spectra for all SSCHP and SSCG models subjected to constant and exponential 
wakes with the same wake decay rates are presented in the
Appendix~\ref{secAP:Spectra} for the reader to consult;
qualitatively, they are similar to the spectra presented in 
Fig.~\ref{fig:SSCConstWakeSpectra}.

%------------------------------------------------------------------------------%
%------------------------------------------------------------------------------%

%------------------------------------------------------------------------------%
\begin{figure*}[t!p]
\includegraphics[width=0.19\linewidth]{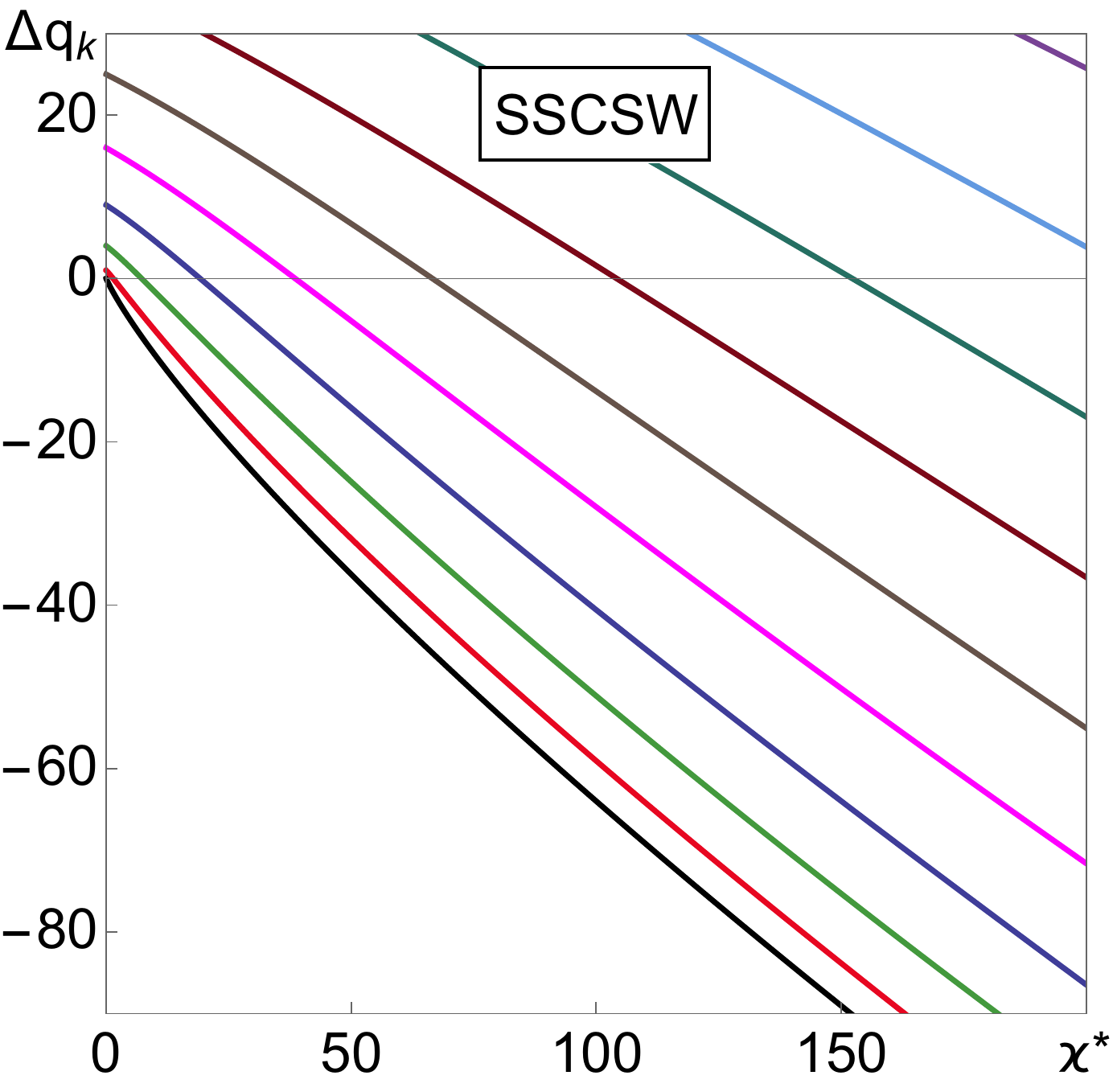}
\includegraphics[width=0.19\linewidth]{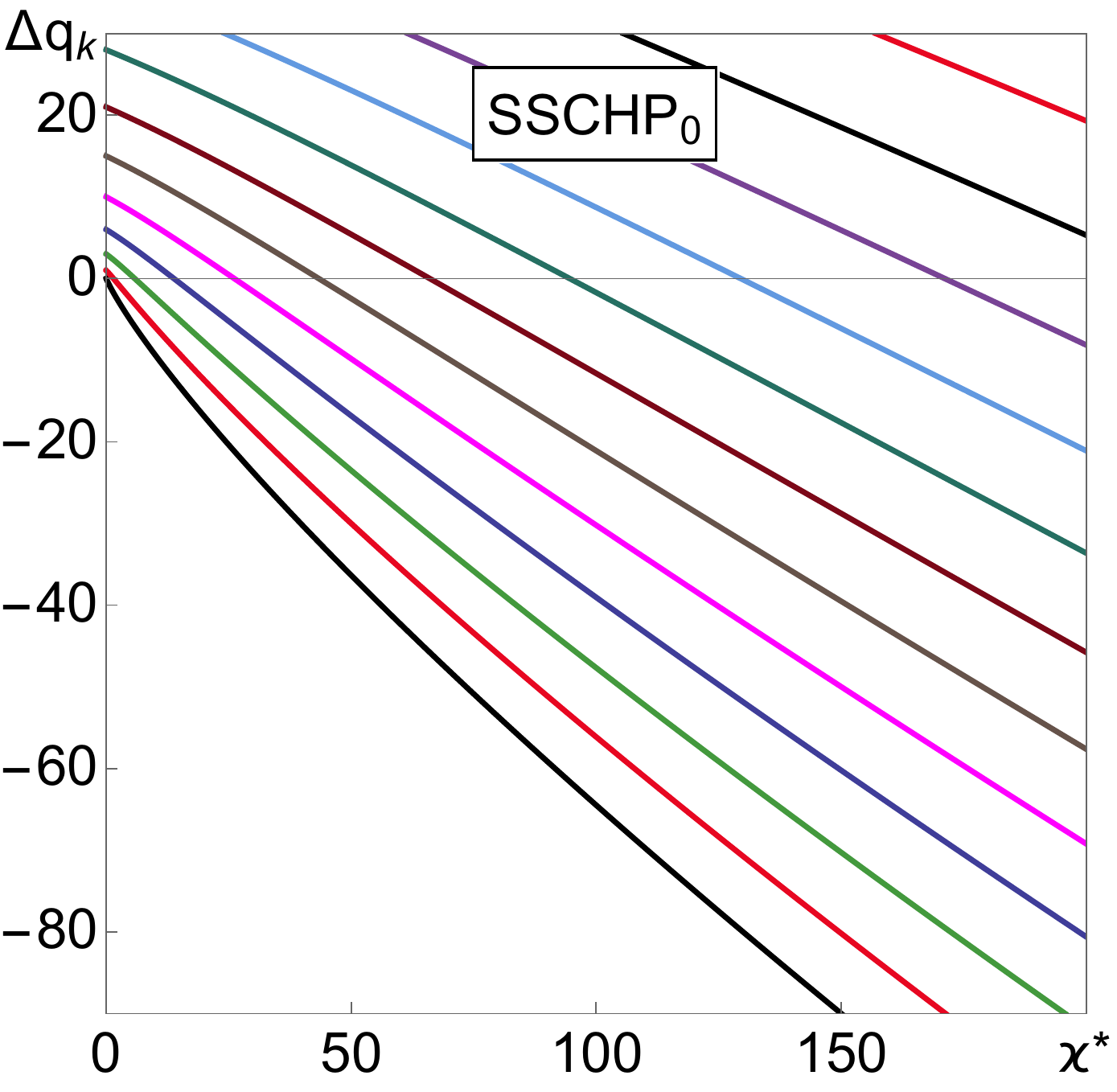}
\includegraphics[width=0.19\linewidth]{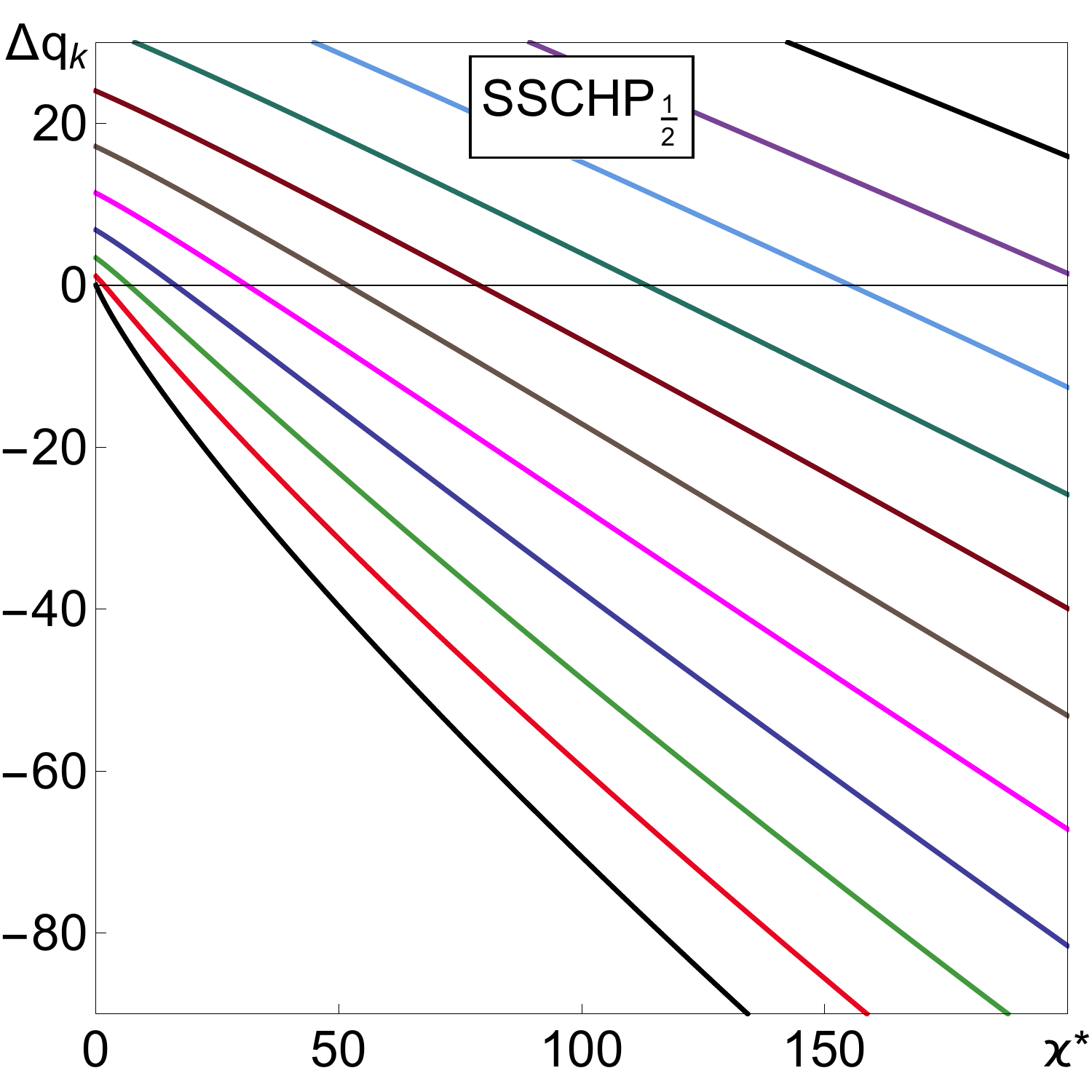}
\includegraphics[width=0.19\linewidth]{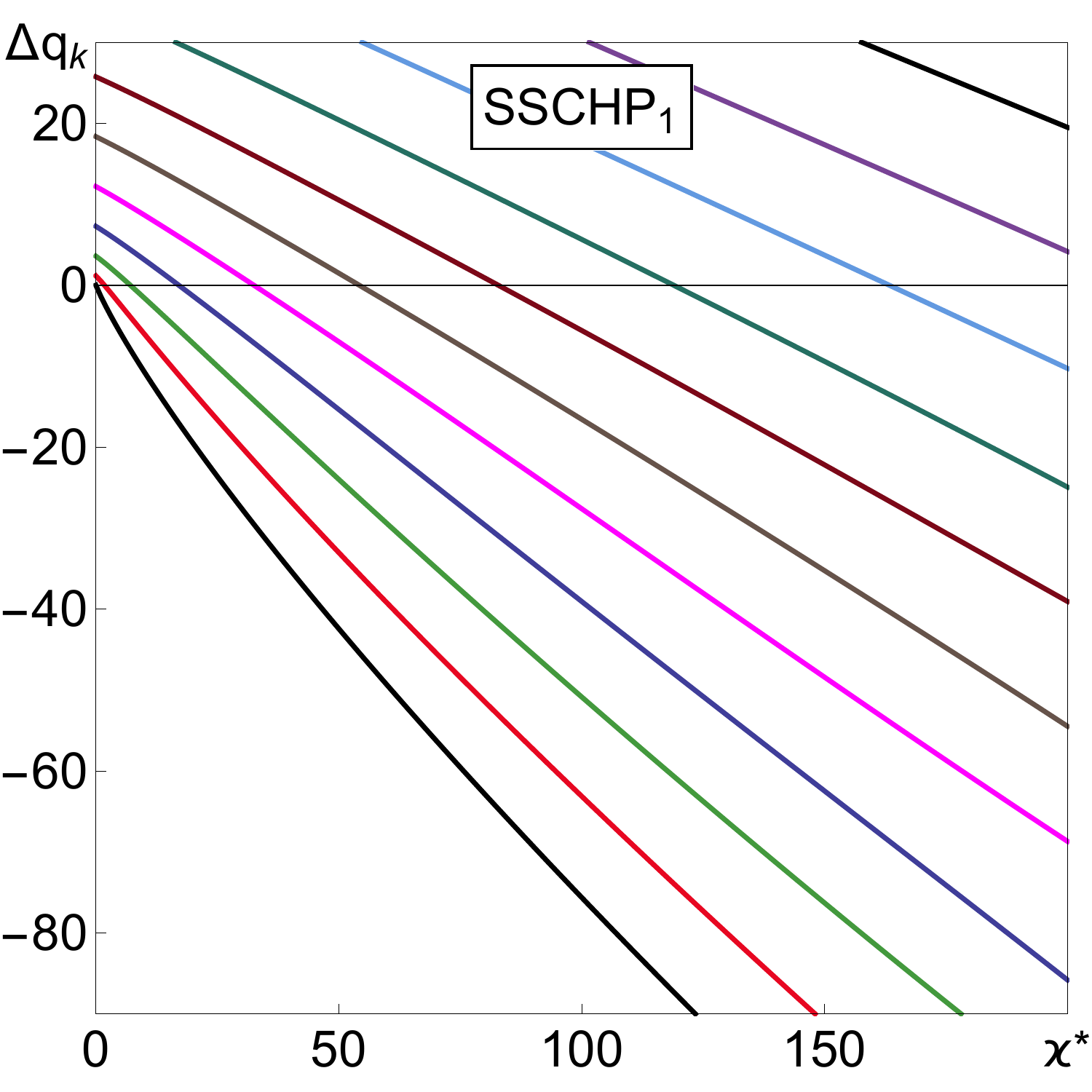}
\includegraphics[width=0.19\linewidth]{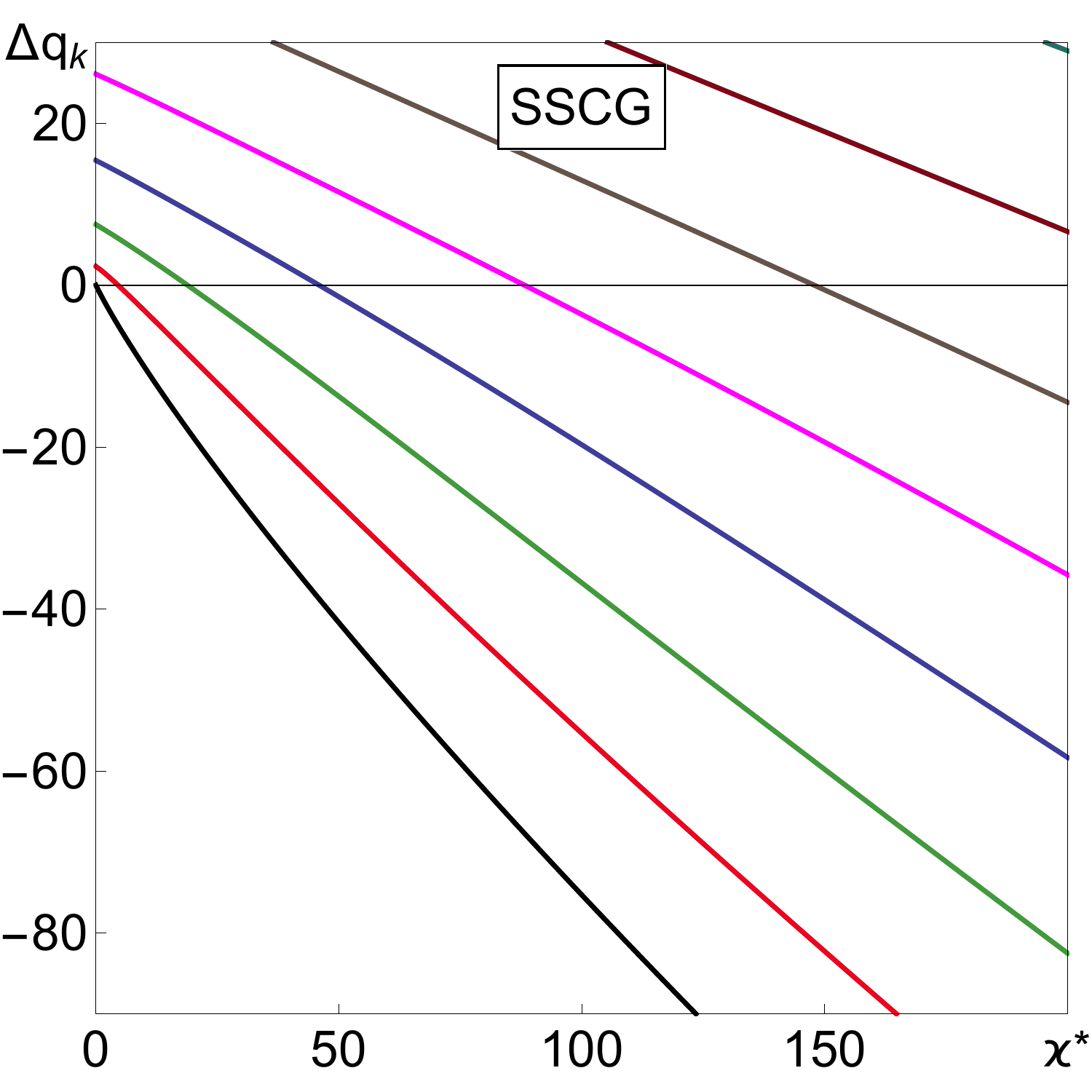}
\caption{\label{fig:SSCResWakeSpectra}
	Spectra for SSC models with resistive wall wake, truncated at 40 modes. 
	No TMCI.
	}
\end{figure*}
%------------------------------------------------------------------------------%

%==============================================================================%
%------------------------------------------------------------------------------%
%==============================================================================%
\subsection{\label{sec:ResWake}Resistive wall wake}

%------------------------------------------------------------------------------%
The SSC models can also be considered with the resistive wall wake function
\begin{equation}
	W(\tau) =
		-W_0/\sqrt{|\tau|}.
\end{equation}
Similar to the exponential and constant wakes, the apparent threshold increases 
without limit as more modes are taken into account, as it can be seen in 
Fig.~\ref{fig:SSCNegWakeConv};
spectra are shown above in Fig~\ref{fig:SSCResWakeSpectra}.

%\newpage
%==============================================================================%
%==============================================================================%
\subsection{\label{sec:StepWake}Step Wake}
%==============================================================================%

%------------------------------------------------------------------------------%
When the wake is shorter than the bunch length, sometimes it is convenient to 
approximate it by a step function
\begin{equation}
W(\tau) =
	W_{0} \left(\text{H}\,[-\tau-\tau_\text{w}]-\text{H}\,[-\tau]\right),
\end{equation}
where $\tau_\text{w} \in (0,1)$ is a characteristic wake length.
While this wake was studied with SW and HP$_0$ models by V.~Balbekov,
we extend the result for Gaussian bunch and SSCHP$_{1/2,1}$ cases.
It was established that, for any value of $\tau_\text{w}$, TMCI vanishes for
all SSC models.
The examples of spectra for $\tau_\text{w} = 0.2,0.4,0.6$ and $0.8$ have been 
included in the Appendix~\ref{secAP:Spectra}.

%------------------------------------------------------------------------------%
\begin{figure*}[h!]
\includegraphics[width=\linewidth]{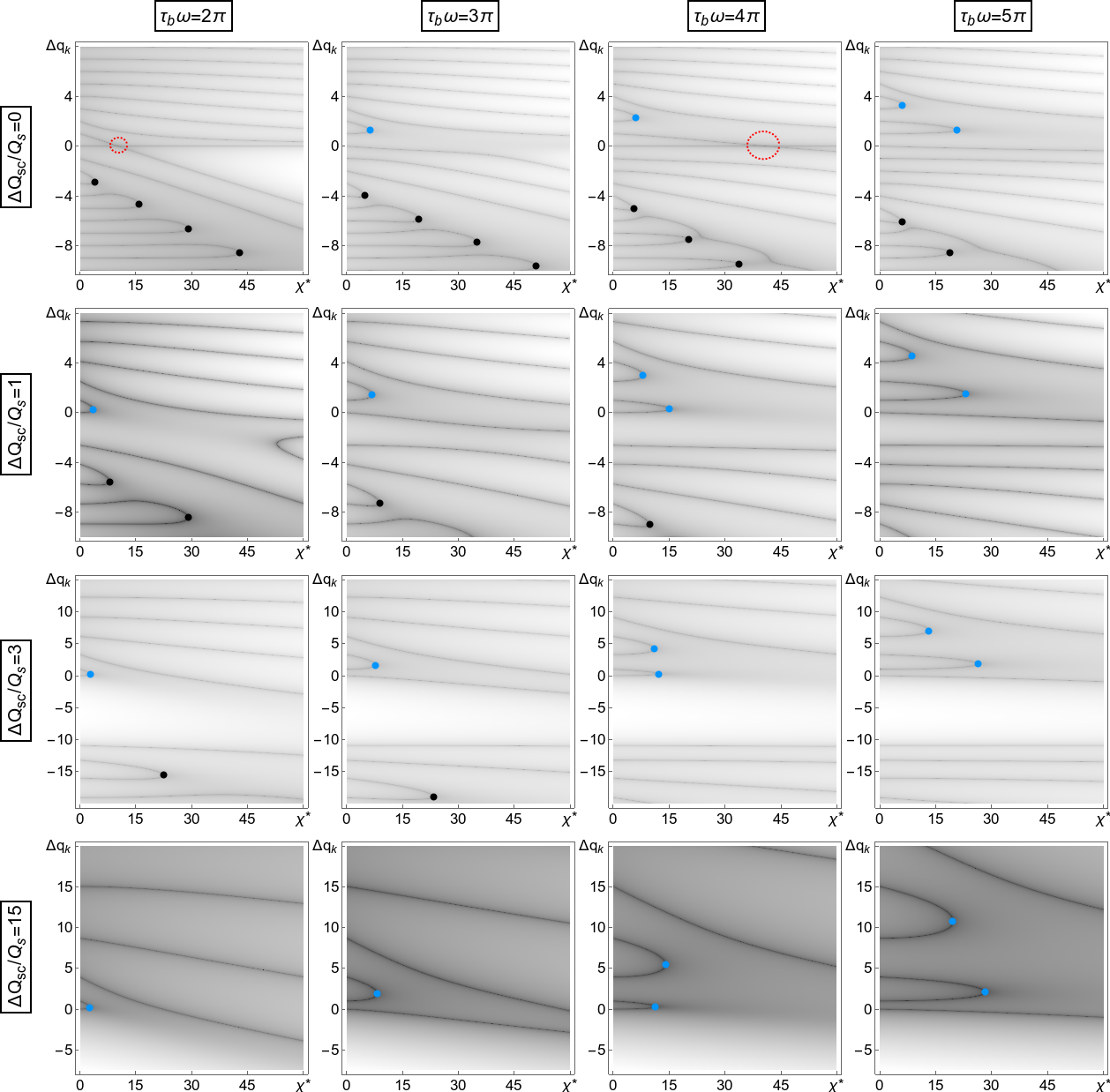}
\includegraphics[width=\linewidth]{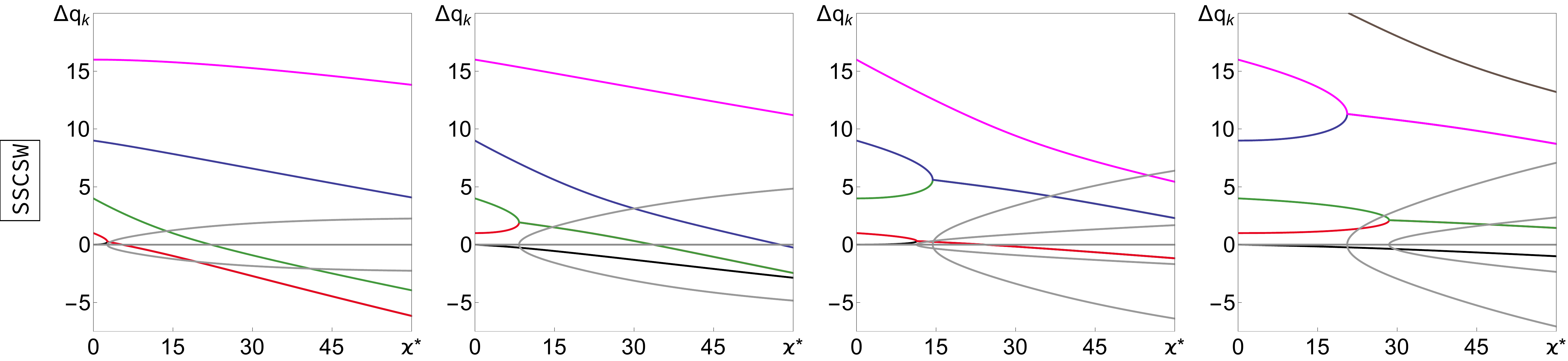}
\caption{\label{fig:CosWakeSpectra}
	Bunch spectra for the ABS model with cosine wake.
	The top four rows correspond to various values of SC parameter
	(0, 1, 3, 15).
	The vanishing TMCI thresholds are shown with black points and 
	non-vanishing with blue.
	The red dashed circles indicate {\it mode crossing}, when modes do not 
	couple.
	The last row is for the SSC limit --- SSCSW model;
	the real and imaginary parts of $\qx$ are shown in colors and in 
	gray respectively.
	The columns correspond to different values of $\omega\,\tb$:
	$2\,\pi$, $3\,\pi$, $4\,\pi$ and $5\,\pi$ from the left to the right 
	respectively.
	}
\end{figure*}
%------------------------------------------------------------------------------%

%==============================================================================%
%------------------------------------------------------------------------------%
%------------------------------------------------------------------------------%
%------------------------------------------------------------------------------%
%==============================================================================%
\section{\label{sec:OscWakes}Oscillating wakes}

%------------------------------------------------------------------------------%
As shown above, the instability for negative wakes takes place only when wake
and space charge tune shifts are comparable,
i.e. the TMCI is of the {\it vanishing} type.
At positive (non-physical) wakes, the instability threshold monotonically 
decreases with the space charge;
the wake moves 0-th tune up, while the space charge deflects the tune of the 
above 1-st mode down, so SC helps the two modes to meet each other.
Since the positive wakes are impossible~\cite{chao1993physics}, they are not 
discussed any more in our paper.
What is possible, though, is a combination of positive and negative wakes, i.e. 
oscillatory wakes.
It can be expected that, at least for some of them, the instability threshold
decreases with the SC, similar to the positive wakes.
In fact, this was already shown for the cosine wake in
Ref.~\cite{burov2009head}:
in contrast with negative wakes, the downward deflection of the cosine wake 
acting on one mode may be more than the deflection of the lower neighboring 
mode causing mode crossing and possible coupling.
Since SC reduces the mode separation, in this case, the instability threshold 
goes down with an increase of the SC tune shift.

%------------------------------------------------------------------------------%
In this section, we consider oscillating wakes of two types:
cosine and sine with variable decay rates.
All of them can be treated within the ABS model, as suggested in 
Ref.~\cite{blaskiewicz1998fast}.
For comparison, results of the complimentary SSC approximation are also provided
below.

%==============================================================================%
%------------------------------------------------------------------------------%
%==============================================================================%
\subsection{\label{sec:CosWake}Cosine wake}

%------------------------------------------------------------------------------%
\begin{figure}[t!]
\includegraphics[width=\linewidth]{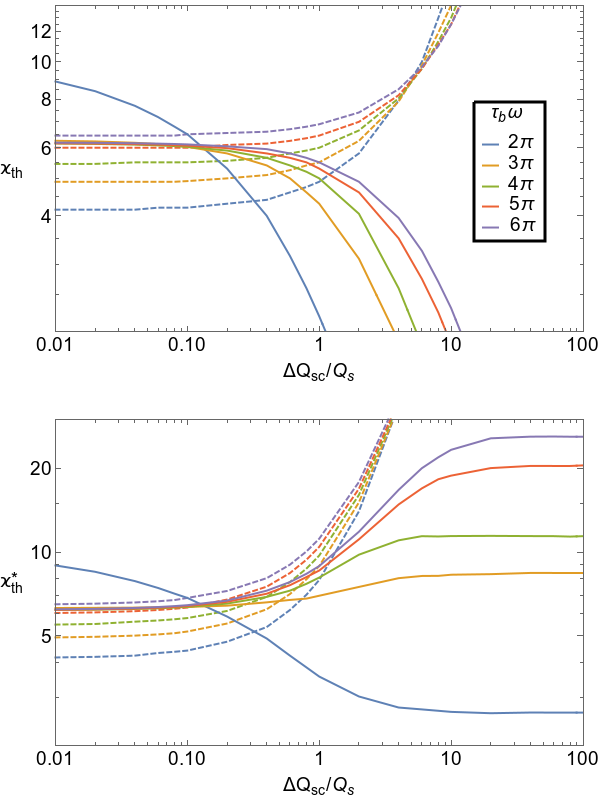}
% {FIGURE_ABCosWakeSummary.pdf}
\caption{\label{fig:ABCosWakeSummary}
	TMCI threshold as a function of space charge for the ABS model and a
	cosine wake.
	The dashed and solid lines correspond to the vanishing and non-vanishing
	TMCIs.
	The top figure shows thresholds in regular units.
	The bottom figure shows the same in terms of the normalized wake 
parameter. 
	}
\end{figure}
%------------------------------------------------------------------------------%

%------------------------------------------------------------------------------%
\begin{figure}[h!]
\includegraphics[width=0.95\linewidth]{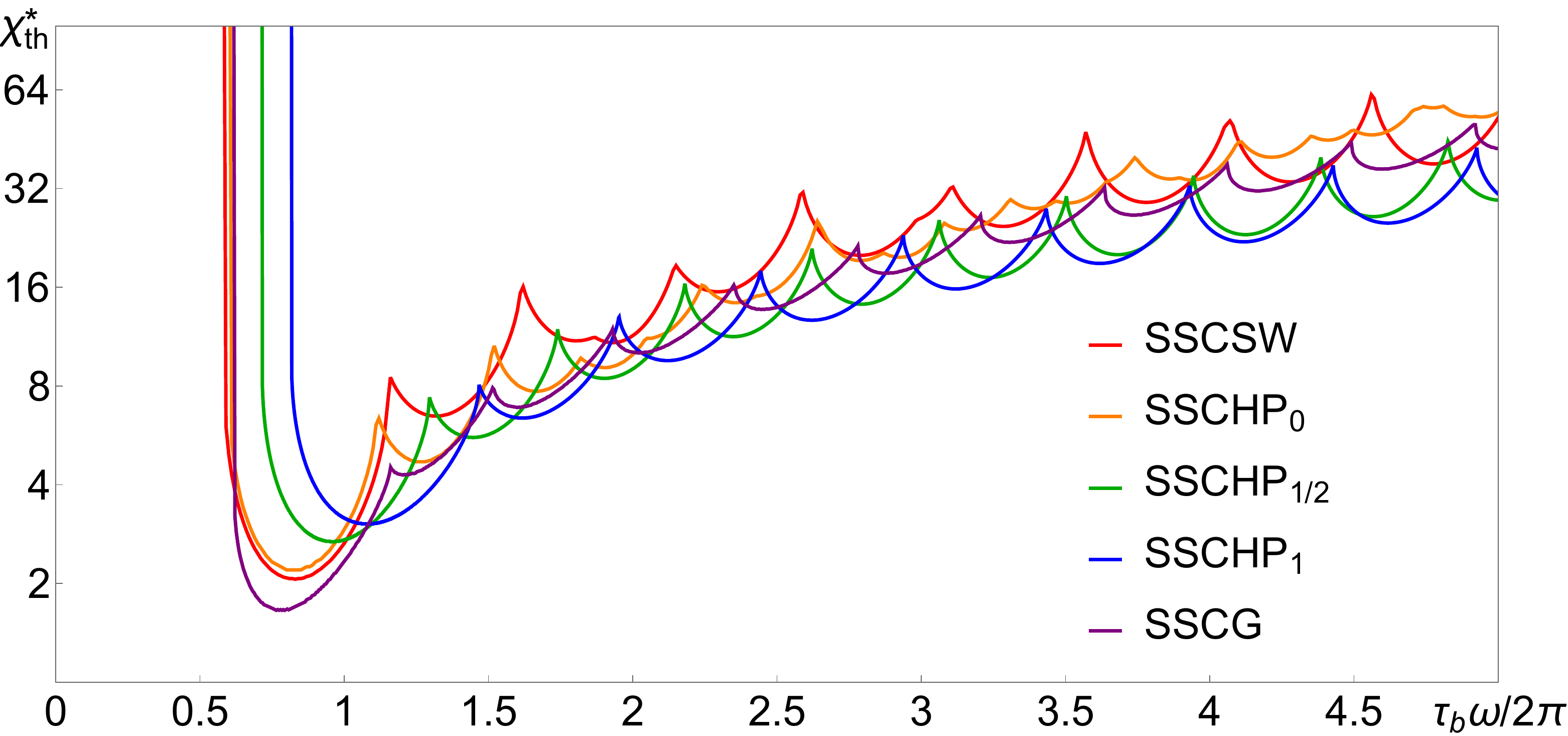}
\caption{\label{fig:SSCCosWakeOmega}
	TMCI threshold for all SSC models with cosine wake as a function of 
	wake phase advance $\omega\,\tb$.
	The vertical lines reflect an absolute threshold with respect to 
	$\omega\,\tb$.
	}
\end{figure}
%------------------------------------------------------------------------------%

%==============================================================================%
%==============================================================================%
\subsubsection{ABS Model}
%==============================================================================%

%------------------------------------------------------------------------------%
For cosine wakes
\begin{equation}
	W(\tau) =
		-W_0 \cos(\omega\,\tau),
\end{equation}
the normalized spectra for different values of the SC tune shift and the wake 
phase advance $\omega\,\tb$ are shown in Fig.~\ref{fig:CosWakeSpectra}.
When the oscillations are sufficiently pronounced
($\omega\,\tb \gtrapprox \pi$), the spectra show that the situation is more 
diverse here than for the negative wakes:
in addition to instabilities in the negative part of the spectra 
(indicated with black points again), there is a mode coupling for positive modes
(blue points).
Figure~\ref{fig:ABCosWakeSummary} summarizes the behavior of the lowest mode 
coupling thresholds for the modes with negative and positive indexes.
While the TMCIs in the negative part of the spectra vanish, as in the case 
of constant-sign wakes, TMCI thresholds for positive modes are going down, as 
was speculated in Ref.~\cite{burov2009head};
as a result, the TMCI threshold is non-monotonic.
On the bottom plot, which repeats the top one in normalized units, one can see 
the saturated TMCI threshold at high SC, which reflects its inverse 
proportionality to the SC parameter.

%==============================================================================%
%==============================================================================%
\subsubsection{SSC Limit}
%==============================================================================%

%------------------------------------------------------------------------------%
In order to summarize the behavior of TMCI in the SSC limit when the bunch 
is subjected to cosine wake, we considered the threshold as a function of 
$\omega\,\tb$ for all models, see Fig.~\ref{fig:SSCCosWakeOmega}.
For smaller values of the wake phase advance, the wake is effectively negative, 
yielding TMCI has a threshold with respect to $\omega\,\tb \approx \pi$, with 
$\tb=3\,\cb$ for SSCG.

%------------------------------------------------------------------------------%
At high SC parameter values, all the models give similar results, with the 
SSCSW model becoming similar to the ABS
(compare the 4-th and 5-th rows in Fig.~\ref{fig:CosWakeSpectra}).
Since the behavior in SSC limit is similar for all cases, the 
corresponding spectra for all SSCHP and SSCG models have been included in the 
Fig~\ref{fig:SSCCosWakeSpectraAPPNDX} of Appendix~\ref{secAP:Spectra}.

%==============================================================================%
%==============================================================================%
\subsubsection{Convergence in SSC approximation}
%==============================================================================%

\begin{figure}[bh!]
\includegraphics[width=0.95\linewidth]{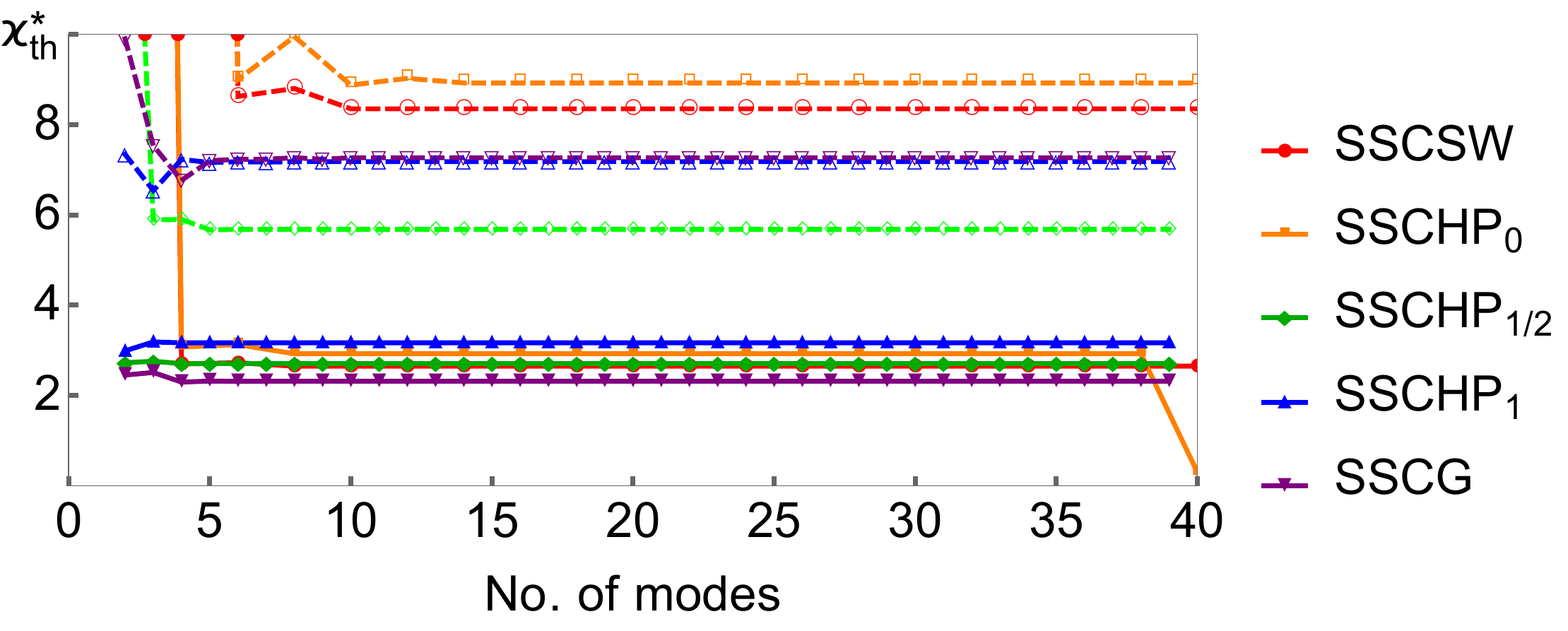}
\caption{\label{fig:SSCCosWakeConv}
	Instability threshold as a function of the number of modes taken into 
	account at the SSC for the cosine wake.
	Solid and dashed lines correspond to different values of wake phase 
	advance, $\omega\,\tb$: $2\,\pi$ and $3\,\pi$ respectively.
	}
\end{figure}

%------------------------------------------------------------------------------%
In contrast to the case of negative wakes, the computed threshold quickly 
converges with the number of modes taken into account.
Examples of convergence of $\chi^*_\text{th}$ for wake phase advance 
$\omega\,\tb=2\,\pi$ and $3\,\pi$ are presented in 
Fig.~\ref{fig:SSCCosWakeConv}.
When the number of modes becomes too large (which is about 30 in our case), 
the SSCHP$_0$ shows a purely numerical instability: 
the threshold computations become unstable and its erroneously computed value 
drops to 0. 

%------------------------------------------------------------------------------%
%------------------------------------------------------------------------------%

%==============================================================================%
%------------------------------------------------------------------------------%
%==============================================================================%
\subsection{\label{sec:SinWake}Sine wake}

\begin{figure*}[ph!]
\includegraphics[width=\linewidth]{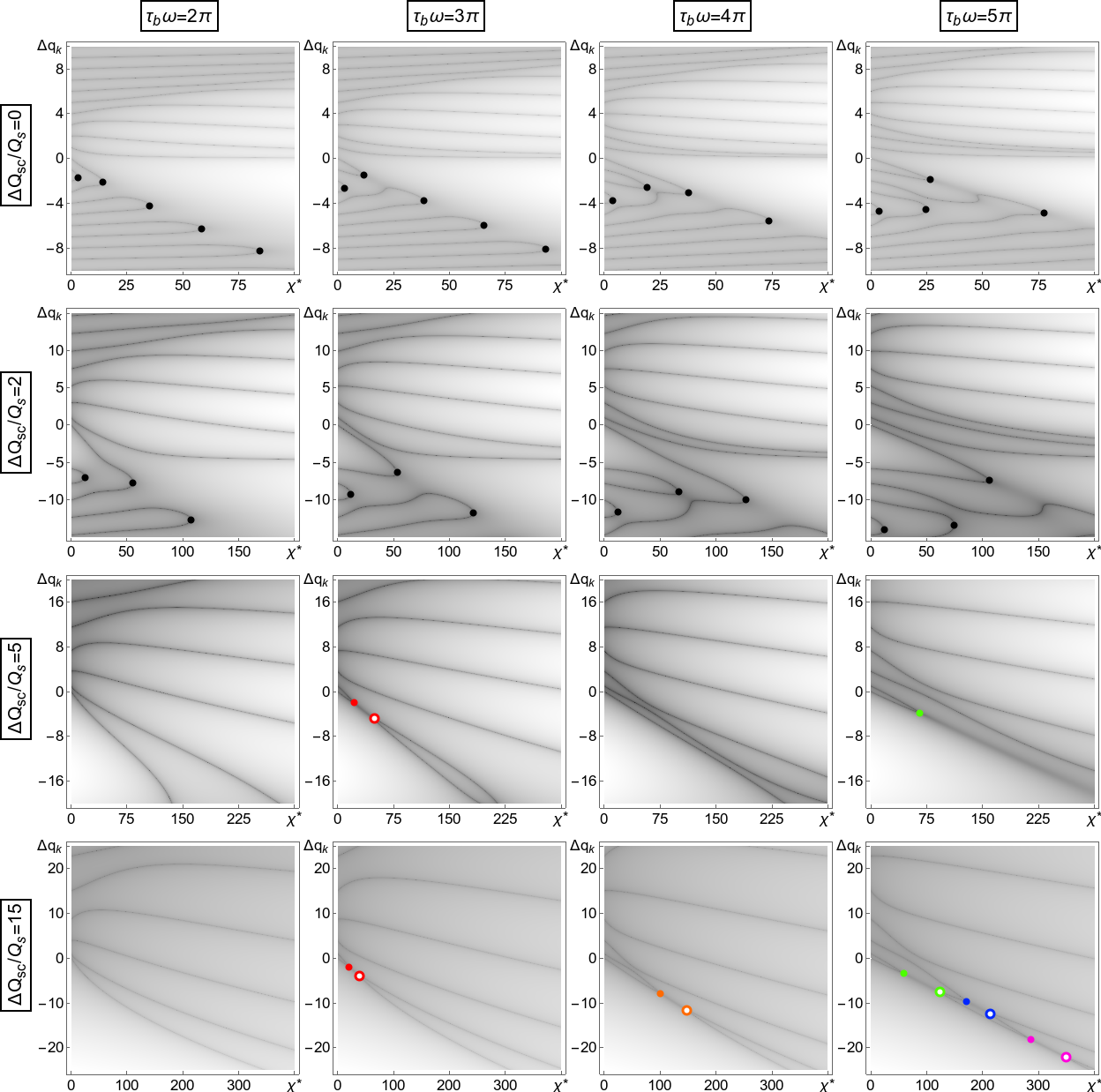}
\includegraphics[width=\linewidth]{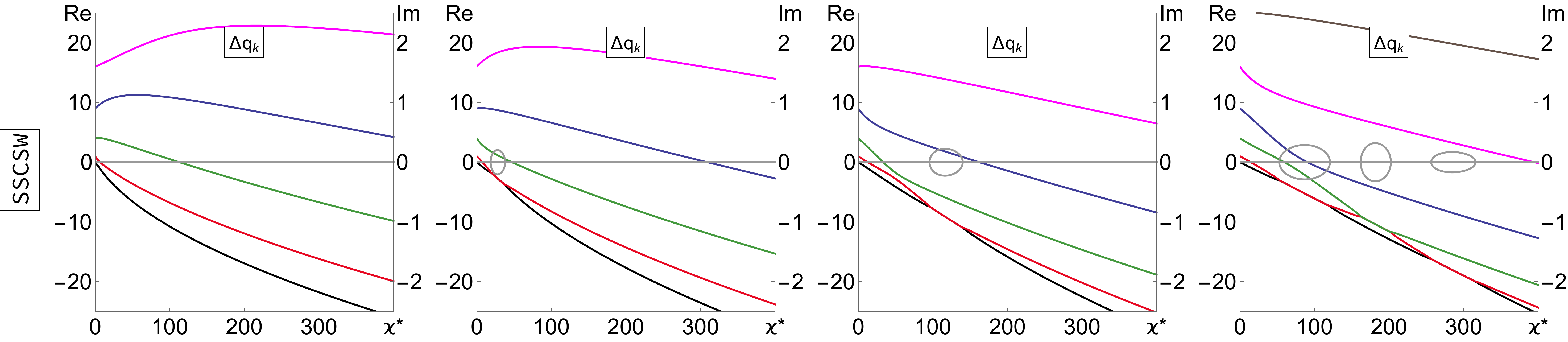}
\caption{\label{fig:SinWakeSpectraABS}
	Bunch spectra for the sine wake.
	The top 4 rows show the ABS spectra for various values of SC 
	(0, 2, 5, 15).
	The vanishing TMCI thresholds are shown with black and non-vanishing 
	using points in colors (subsequent decoupling for each TMCI is shown by 
	the point of the same color with an annulus).
	The last row is for the SSCSW model;
	the real and imaginary parts of $\qx$ are shown in colors and in 
	gray respectively (note the different scaling for real and imaginary 
	parts).
	The columns correspond to the different values of $\omega\,\tb$
	($2\,\pi,3\,\pi,4\,\pi,5\,\pi$).
	}
\end{figure*}

\begin{figure*}[ph!]
\includegraphics[width=\linewidth]{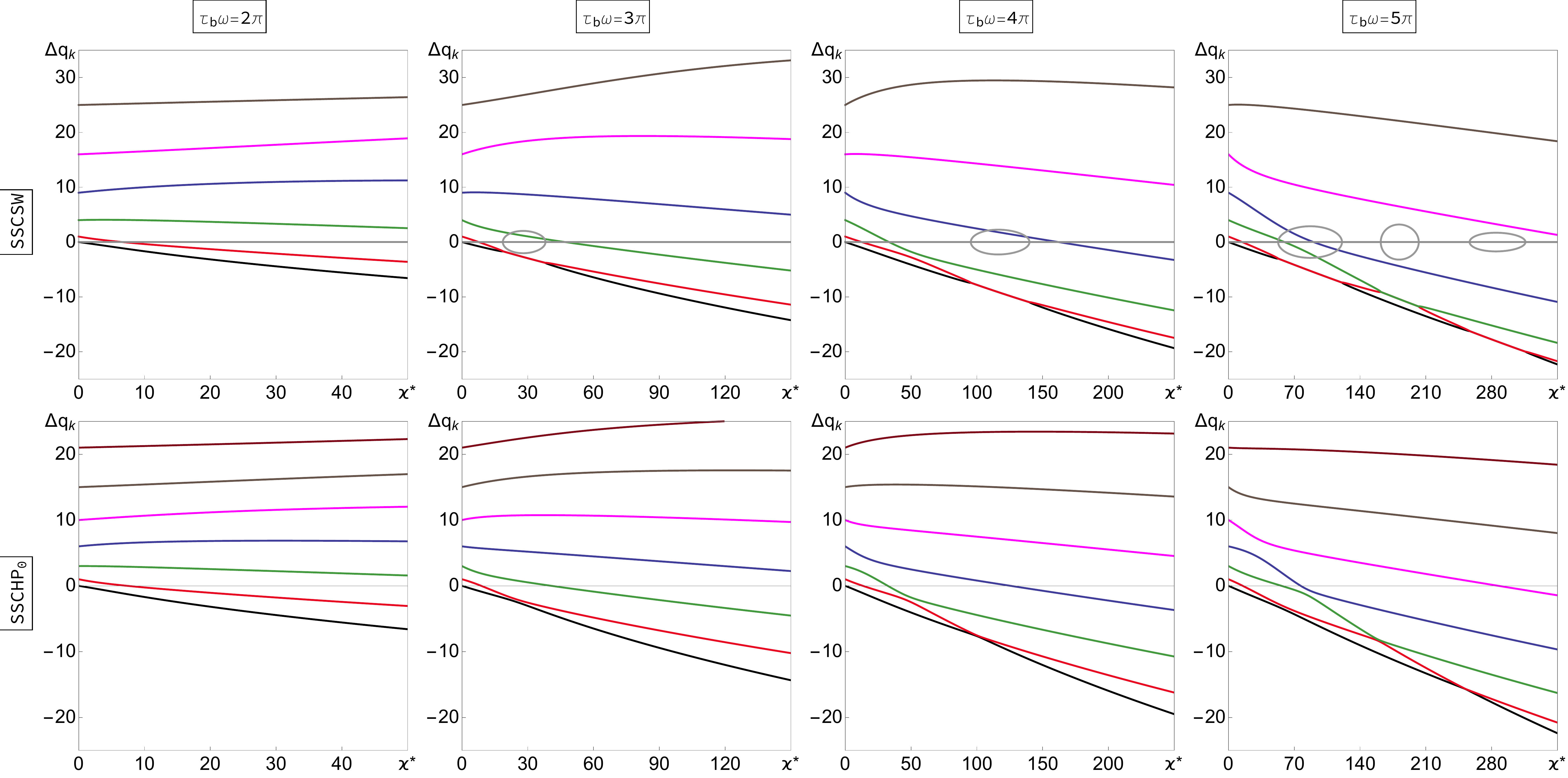}
\includegraphics[width=\linewidth]{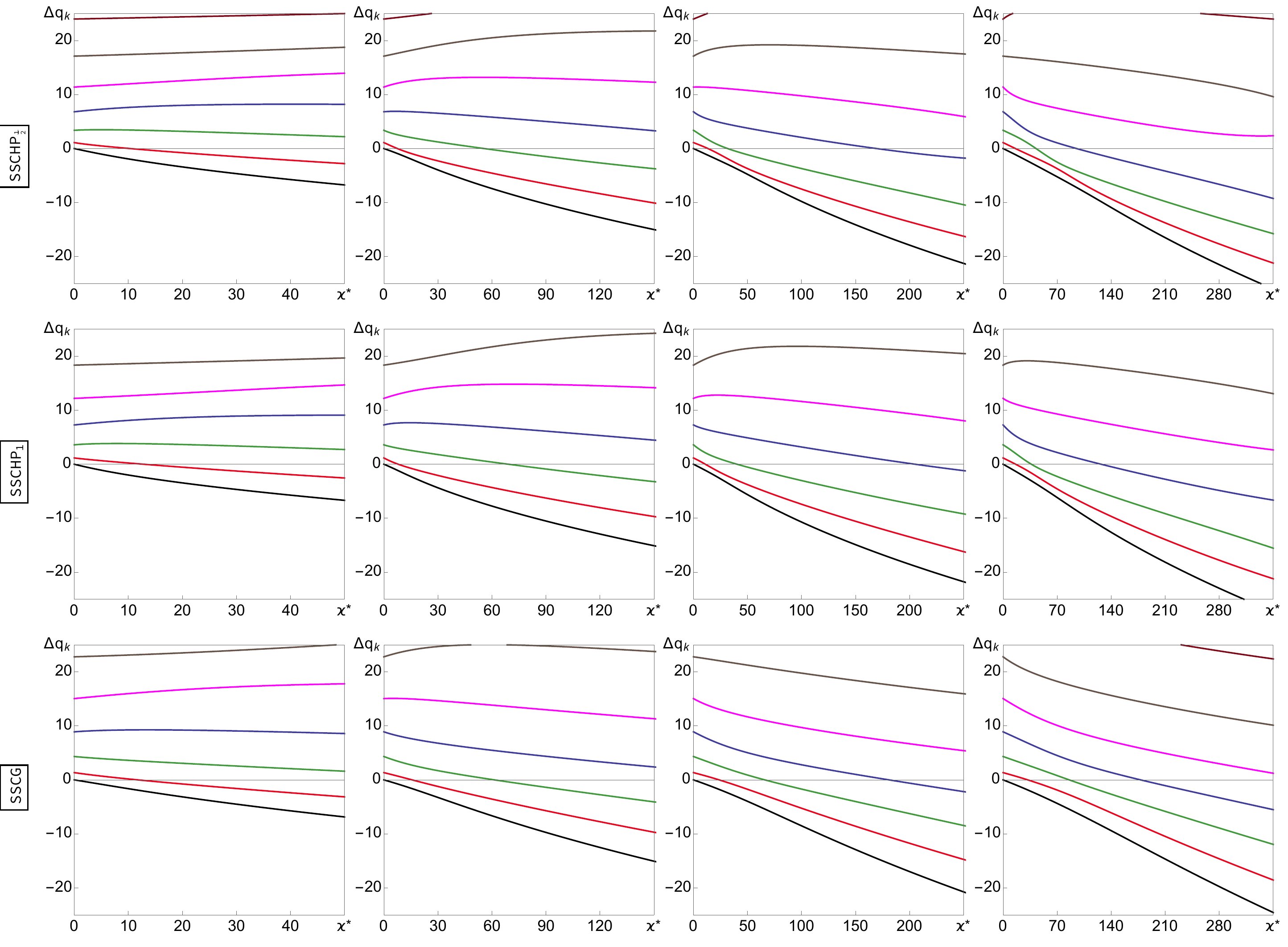}
\caption{\label{fig:SinWakeSpectraSSC}
	Bunch spectra for the sine wake with all SSC models.
	The real and imaginary parts of $\qx$ are shown in colors and in 
	gray respectively (the scaling for the real and imaginary parts is the 
	same as in Fig.~\ref{fig:SinWakeSpectraABS}).
	The columns correspond to the different values of $\omega\,\tb$
	($2\,\pi,3\,\pi,4\,\pi,5\,\pi$).
	The instability is observed only for SSCSW model.
	SSCHP$_0$ spectra show mode crossing without coupling.
	}
\end{figure*}

%------------------------------------------------------------------------------%
The sine wake
\begin{equation}
	W(\tau) = W_0\,\sin(\omega\,\tau),
\end{equation}
is considered for the same models as for the cosine wake and the resulting 
spectra are shown in Figs.~\ref{fig:SinWakeSpectraABS}
and~\ref{fig:SinWakeSpectraSSC}.

%------------------------------------------------------------------------------%
%------------------------------------------------------------------------------%

%------------------------------------------------------------------------------%
%------------------------------------------------------------------------------%

%------------------------------------------------------------------------------%
\begin{figure}[b!]
\includegraphics[width=\linewidth]{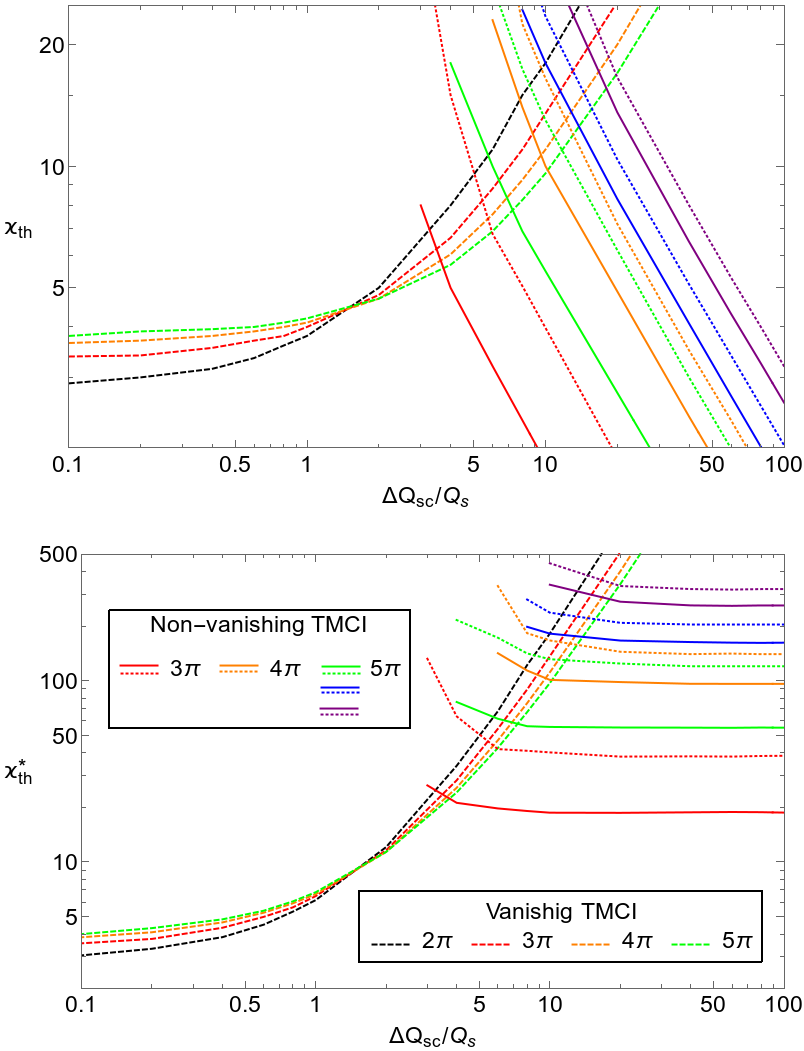}
\caption{\label{fig:SinWakeABSummary}
	TMCI threshold in terms of the usual (top) and normalized (bottom) wake 
	parameters versus the SC parameter for the ABS model and sine wake.
	The dashed lines correspond to the vanishing TMCI in a negative part of 
	the spectra.
	The solid and dotted lines represent first couplings and first 
	decouplings for positive modes.
	}
\end{figure}
%------------------------------------------------------------------------------%

%==============================================================================%
%==============================================================================%
\subsubsection{ABS Model}
%==============================================================================%

%------------------------------------------------------------------------------%
In the ABS model, the behavior of the negative modes looks similar to the cases 
of the cosine and negative wakes.
In contrast, for the positive modes, there is no instability when the SC is 
zero, independently of the wake phase advance $\omega \tb$
(first row in Fig.~\ref{fig:SinWakeSpectraABS}).
However, when the SC is increased, a cascade of mode couplings and decouplings 
appears (3rd and 4th rows in Fig.~\ref{fig:SinWakeSpectraABS}).
When $\omega\,\tb = 2\,\pi$, the threshold increases with the SC parameter, 
disappearing at the SSC case. 
The cases of $\omega\,\tb=3\,\pi,4\,\pi$ show a single mode coupling followed 
by decoupling;
the case $\omega\,\tb=5\,\pi$ has a cascade of three couplings with subsequent 
decouplings.
For wakes with even more oscillations per bunch, more complicated 
coupling-decoupling cascades were observed (not shown in the figure).
Similar to the cosine wake, Fig.~\ref{fig:SinWakeABSummary} summarizes the 
behavior of the lowest TMCI thresholds for both negative and positive modes. 
While all instabilities in the positive part of the spectrum decouple at higher 
wake parameter, the values of these coupling and decoupling intensities 
decrease inversely proportional to the SC parameter.

%------------------------------------------------------------------------------%
\begin{figure}[t!]
\includegraphics[width=0.8\linewidth]{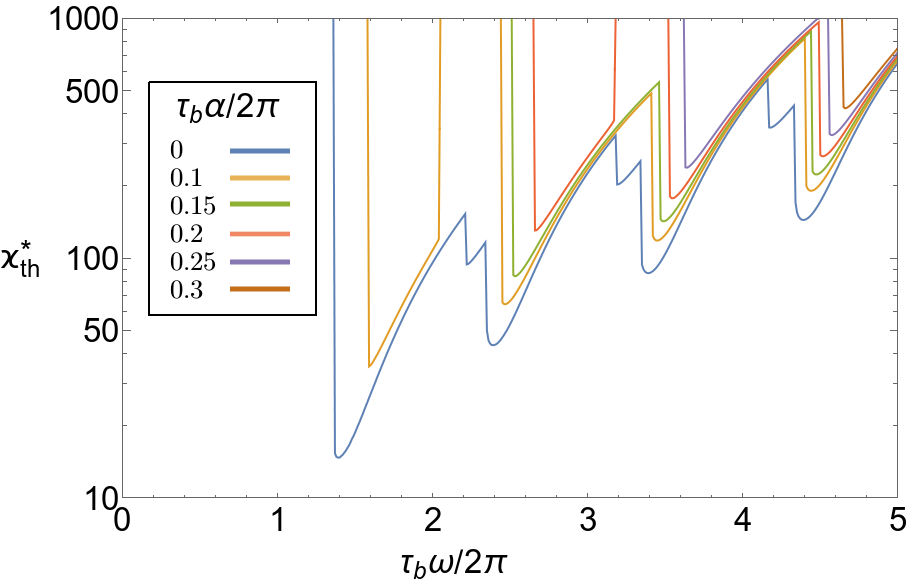}
\caption{\label{fig:SSCSinWakeOmega}
	TMCI threshold for SSCSW model with resonator wake as a 
	function of wake phase advance $\omega\,\tb$.
	The different curves correspond to the different values of the wake 
	decay rate, $\alpha\,\tb$.
	The blue curve with $\alpha\,\tb = 0$ is for the sine wake.
	The vertical lines reflect an absolute threshold with respect to 
	$\omega\,\tb$.
	}
\end{figure}
%------------------------------------------------------------------------------%

%==============================================================================%
%==============================================================================%
\subsubsection{SSC limit}
%==============================================================================%

%------------------------------------------------------------------------------%
Figure~\ref{fig:SinWakeSpectraSSC} shows the spectra for all SSC models.
The first row is the spectra for SSCSW which are in agreement with the ABS ones 
when the SC parameter is high enough (as expected).
The spectra for SSCHP$_{0}$ is shown in the second row.
Although it may look similar to both ABS and SSCSW models, they are different 
because the SSCHP$_{0}$ modes never couple.
Instead, they either simply cross or approach-divert with respect to each other 
without coupling.
It looks like there is an unknown reason forbidding the instability, and not 
only for this case.

%------------------------------------------------------------------------------%
Indeed, for more realistic distributions, i.e. the SSCHP$_{1/2,1}$ and SSCG,
there are no instabilities as well
(last three rows in Fig.~\ref{fig:SinWakeSpectraSSC}).
For these cases, the modes never cross each other.
Thus, we conclude that there is no TMCI for these more realistic bunch 
distributions subjected to a sine wake.

%------------------------------------------------------------------------------%
The behavior of the non-vanishing TMCI as a function of $\omega\,\tb$ for SSCSW 
is shown in Fig.~\ref{fig:SSCSinWakeOmega}.
Similar to the cosine wake, the instability is impossible when
the oscillations are not sufficiently pronounced,
$\omega\,\tb \lessapprox 3\,\pi$.
Otherwise the  instability threshold is a non-monotonic function of the 
wake phase advance $\omega\,\tb$.

%==============================================================================%
%------------------------------------------------------------------------------%
%==============================================================================%
\subsection{\label{sec:ROsc}Resonator wake}

%------------------------------------------------------------------------------%
Finally, we consider the spectrum for a decaying sine wake:
\begin{equation}
	W = W_0\,\sin(\omega\,\tau)\,e^{\alpha\,\tau};
\end{equation}
conventionally, this wake function is referred as resonator.

%------------------------------------------------------------------------------%
The result for the SSCSW model is shown in Fig.~\ref{fig:SSCSinWakeOmega}.
There is no surprise that the instability threshold increases with $\alpha$ for 
fixed $\omega$.
When a certain threshold with respect to the decay parameter $\alpha$ is 
reached, the instability vanishes: strong enough suppression of the oscillating 
part by the exponential decay makes the wake effectively negative.
Thus, the TMCI vanishes in certain zones of the wake decay and oscillation
parameters $\alpha \tb$ and $\omega \tb$.

%------------------------------------------------------------------------------%
The ABS model confirms this result: for fixed $\omega$, when a threshold with 
respect to $\alpha$ is reached, instabilities in the positive part of the 
spectrum do not appear, even for large values of space charge.
The only instabilities are for the negative modes, vanishing with an increment 
of SC, as expected.
A particular example for CERN SPS ring is discussed in next
Subsection~\ref{sec:SPS}.

%------------------------------------------------------------------------------%
In the SSCG and all SSCHP models there is no instability as was in the case of 
the sine wake;
the mode crossing observed in the situation of the SSCHP$_0$ case turns into an
approach-divert behavior and the modes separate when $\alpha$ is increased.

%------------------------------------------------------------------------------%
As an example, we compare the SSC theory results with macroparticle simulations
of Ref.~\cite{blaskiewicz2012comparing}.
Two different bunch distributions, HP$_0$ and HP$_{7/2}$, were studied with 
constant and two resonator wake potentials:
\[
\begin{array}{l}
	W_1(\tau) = W_0\,\sin(12\,\tau)\,e^{1.2\,\tau},	\\[0.2cm]
	W_2(\tau) = W_0\,\sin(24\,\tau)\,e^{2.4\,\tau},
\end{array}
\]
with $\tau$ in units of $\tb$.
According to Fig.~3 from~\cite{blaskiewicz2012comparing}, the dependence of the 
threshold wake strength as a function of the space charge tune shift is a 
non-monotonic function.
For some cases (HP$_{7/2}$ with constant and both HP$_{0,7/2}$ with $W_1$), the 
SC stabilizes the beam for $\Qsc \lessapprox 2\,\Qs$, but then thresholds are 
getting smaller than without SC, when the last one is increased.
For HP$_0$ with constant wake, the simulations demonstrated stabilization of 
TMCI with SC with a sign of turning over at larger values of SC.
Finally, for HP$_{7/2}$ with $W_2$ wake, the TMCI threshold monotonically 
decreases, showing no improvement with SC.

%------------------------------------------------------------------------------%
\begin{figure}[t!]
\includegraphics[width=\linewidth]{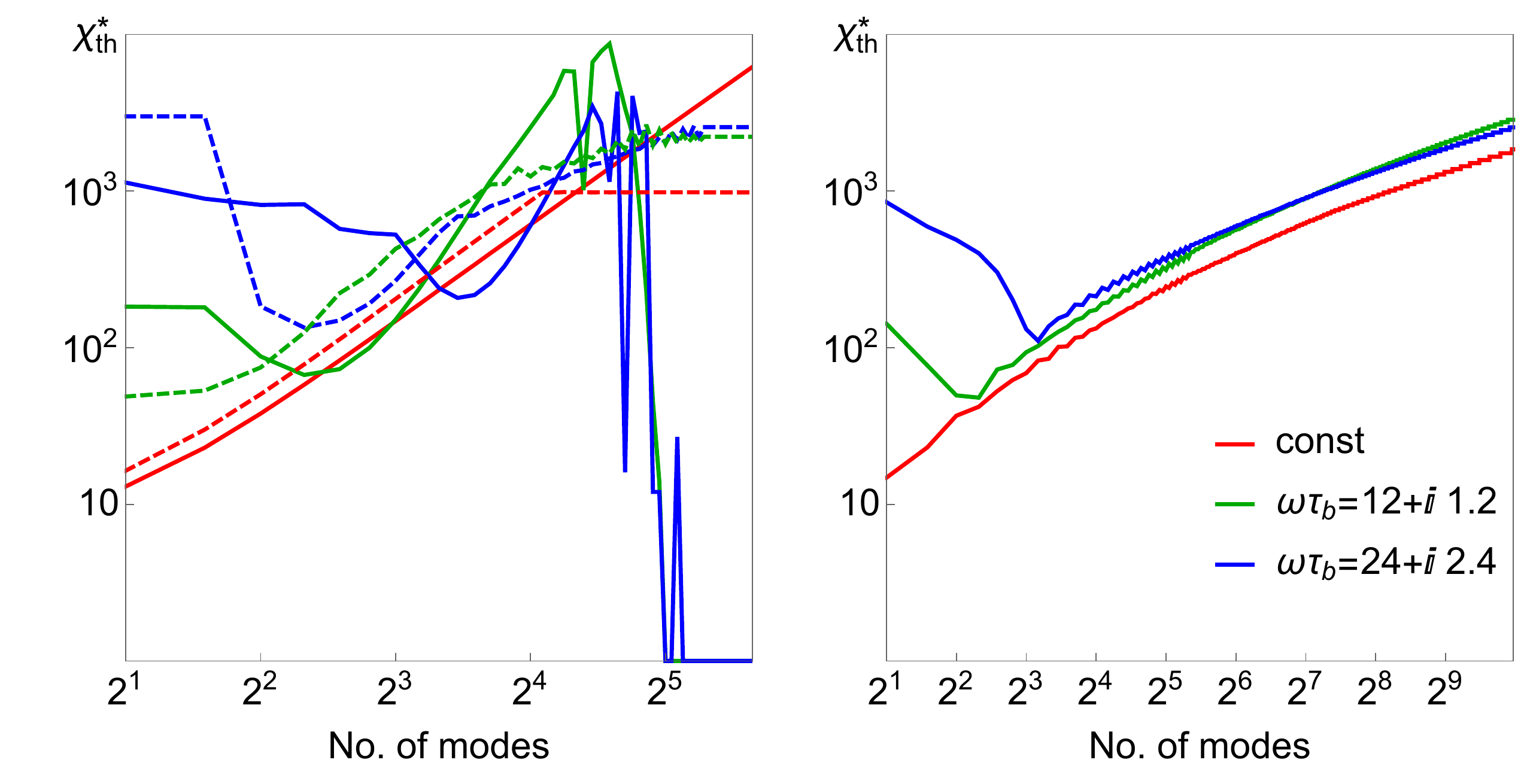}
\caption{\label{fig:MBtest}
	TMCI threshold as a function of the number of modes taken into account
	for SSCHP$_0$ (solid lines) and SSCHP$_{7/2}$ (dashed lines) models
	with wake functions from~\cite{blaskiewicz2012comparing}.
	The saturation with number of modes for SSCHP$_{7/2}$ case is determined
	by precision in calculation of matrix elements.
	For SSCHP$_0$ with resonator wakes there is the numerical instability
	of threshold when the number of modes $\approx 20$.
	Right plot is for SSCHP$_0$ with the same wakes but using expansion 
	in SSCSW basis: no saturation nor numerical instability is observed.
	}
\end{figure}
%------------------------------------------------------------------------------%

%------------------------------------------------------------------------------%
The macroparticle simulations of~\cite{blaskiewicz2012comparing} can be 
compared 
with SSC results shown in Fig.~\ref{fig:MBtest}.
The left figure illustrates the instability threshold as a function of the
number of modes taken into account for a boxcar SSCHP$_0$ and smooth
SSCHP$_{7/2}$ models with the same wake potentials.
Few features of this figure can be noted.
First, for both distributions, the calculated threshold goes down with the 
number of modes until the most resonant mode is added into analysis.
Second, numerical effects for the two distributions are different:
for the smooth SSCHP$_{7/2}$ one can see the saturation with the number of 
modes, while for the boxcar SSCHP$_{0}$ the threshold suddenly drops down after 
the monotonic growth.
In the smooth case, the wake matrix elements were computed using numerical 
integration and one should not expect the correct result when
$\epsilon\,\chi^* \approx 1$, which is about the distance between two closest 
eigenvalues of $\nu_{lm}$, and, $\epsilon$ is the average numerical error for
the element of $\mW$.
In the boxcar case, the matrix elements are given by the analytical expressions
containing multiple summations (see Appendix~\ref{secAP:WlmCONST} for details).
The available algorithm evaluating these sums is numerically unstable for a
sufficiently large number $l+m$;
as a result, the threshold erroneously tends to zero.

%------------------------------------------------------------------------------%
The last effect can be suppressed by using the decomposition of the SSCHP$_0$
modes using SSCSW basis.
Looking for the solution of the Eq.~(\ref{math:Woperator}) in the form
\begin{equation}
\label{math:SWDecomp}
	\Yk^{\text{HP}_0}(\tau) =
		\sum_{i=0}^{\infty} \mathbf{B}^{(k)}_i Y_i^\text{sw}(\tau)
\end{equation}
where SSCSW harmonics (\ref{math:YkSW}) are used instead of Legendre polynomials
(\ref{math:YkHP0}), the eigenvalue problem~(\ref{math:EVProb}) becomes
\begin{equation}
\label{math:EVProbSW}
\mathbf{M}\cdot\mathbf{B}^{(k)} = \qx \mathbf{B}^{(k)},
\qquad
\mathbf{M}_{lm} = \mL + \kappa\,\mW^\text{sw},
\end{equation}
where $\mW^\text{sw}$ is the wake matrix elements computed for SSCSW model, 
and, 
$\Lambda_{lm}$ is the matrix of SSCHP$_0$ harmonics in SSCSW basis
\[
\mL = \left\{
\begin{array}{ll}
	\ds -\frac{(-1)^\frac{3\,l+m}{2}4\,l^2m^2}{(l^2-m^2)^2},	&
	\text{if }l+m\text{ is even},\,l \neq m			\\[0.35cm]
	\ds \frac{1-\delta_{l,0}}{4} + \frac{\pi\,l^2}{12},		&
	\text{if }l=m						\\[0.35cm]
	0,						& \text{otherwise}.
\end{array}
\right.
\]
The result for the same wake potentials is shown in the right plot of 
Fig.~\ref{fig:MBtest}.
The convergence is slower, since we expanding Legendre polynomials using the 
trigonometric series, but matrix elements become more stable and significantly
larger number of modes can be used in verifying the convergence.
As one can see, no sign of numerical instability or saturation is present.

%------------------------------------------------------------------------------%
Comparing the results of macroparticle simulations with the SSC theory, it 
should be noticed the methodical advantage of the last, provided it is 
applicable:
there is only one numerical parameter --- the number of modes, allowing a 
reliable convergence check, while in the first case there are many numerical 
parameters --- a number of macroparticles, number of time steps per betatron 
oscillation, numerical size of delta functions, spacial grid size and the 
simulation time.
Convergence checks with each of them are required, which cannot be an easy 
task. 
Since only one of these checks was reported in~\cite{blaskiewicz2012comparing}, 
a poor convergence with its other numerical parameters could be suspected as a 
reason for the disagreement between Ref.~\cite{blaskiewicz2012comparing} and 
this paper.

%==============================================================================%
%------------------------------------------------------------------------------%
%==============================================================================%
\subsection{\label{sec:SPS}CERN SPS}

%------------------------------------------------------------------------------%
In this last section we will consider the behavior of TMCI for CERN SPS using
ABS model, allowing computations for arbitrary SC parameter.
The wake field of the ring is modeled as a broadband resonator, 
Ref.~\cite{quatraro2010effects},
\begin{equation}
	W(\tau) = Z_t\,\frac{\omega_r^2}{\hat\omega\,Q_r}\,
	\sin(\hat\omega\,\tau)\,
	e^{\alpha \tau /2}, \;\; \tau<0,
	\label{BBWake}
\end{equation}
with
\begin{equation}
	\hat\omega = \sqrt{\omega_r^2-\alpha^2},
\end{equation}
\begin{equation}
	\alpha = \omega_r/(2\,Q_r),
\end{equation}
$\omega_r = 1.3\times2\pi$~GHz,
$Q_r=1$ and
$Z_t = 5$ $\text{M}\Omega/\text{m}$.
For this model, the presumably Gaussian bunch with rms length $\cb = 30$ cm is 
substituted by a boxcar ABS bunch with the total length $\tb = 3\,\cb$.

\begin{figure}[h!]
\includegraphics[width=\linewidth]{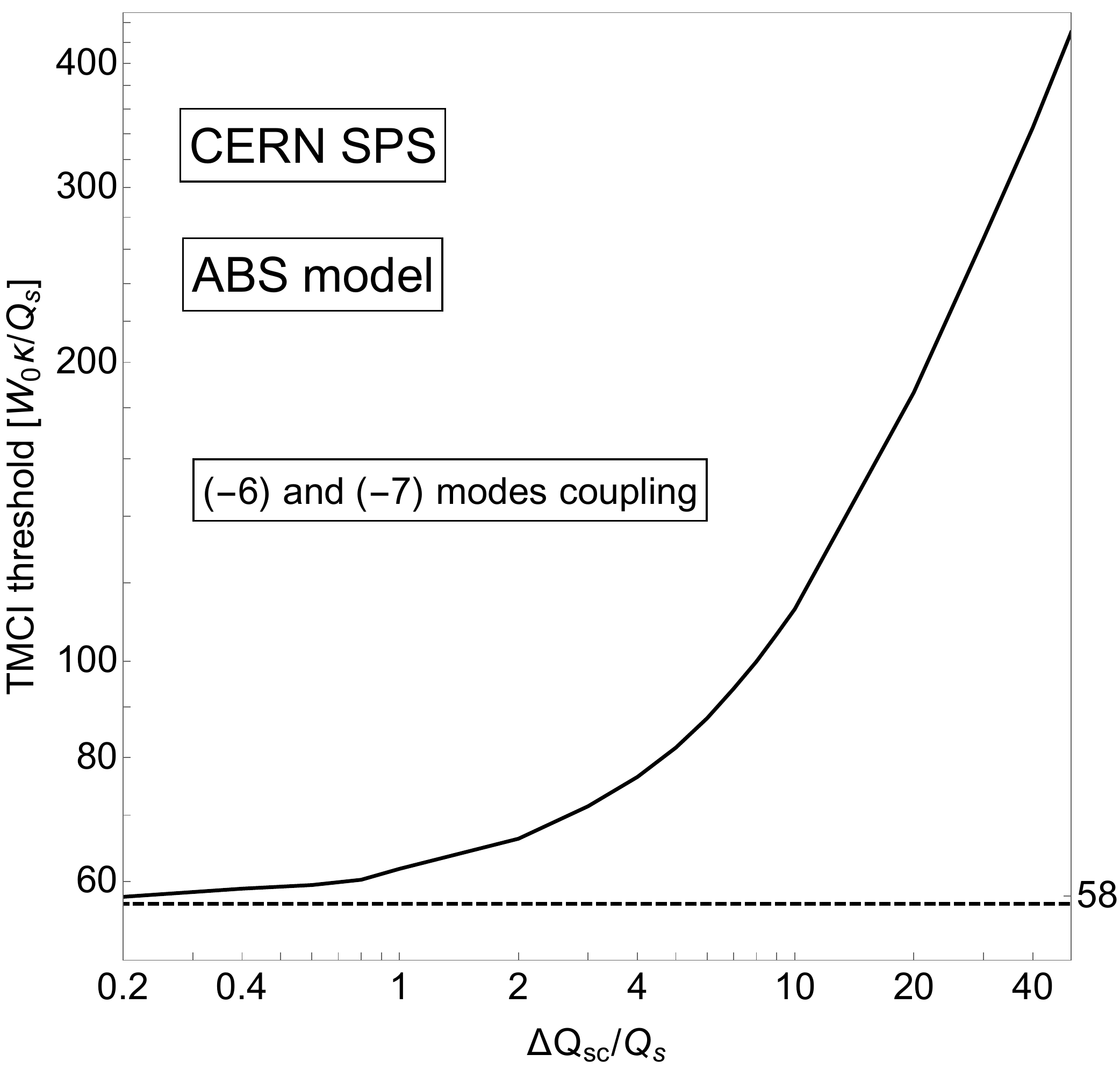}
\caption{\label{fig:SPS}
	The lowest TMCI threshold as a function of space charge for CERN SPS 
	ring (ABS model).
	Dashed line shows the value of threshold at zero $\Qsc$.
	}
\end{figure}

\begin{figure}[h!]
\includegraphics[width=\linewidth]{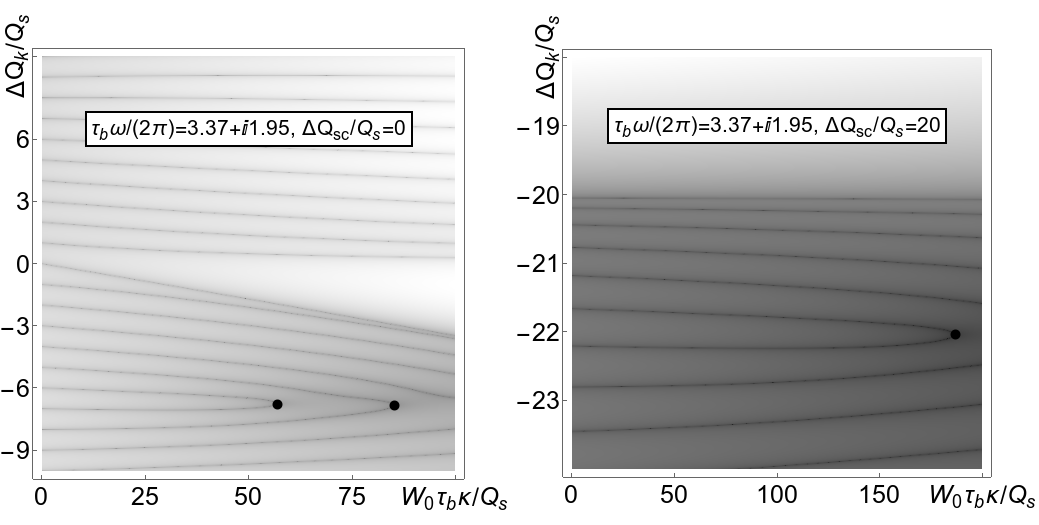}
\caption{\label{fig:SPSSpectra}
	The bunch spectra for the CERN SPS ring in ABS model with broadband 
	resonator wake.
	The vanishing TMCI thresholds are illustrated with black points.
	Left plot shows first 20 modes ($k=-10,\ldots,9$) at no SC.
	Right plot shows only negative part of spectra ($k=-10,\ldots,-1$)
	at $\Qsc/\Qs = 20$;
	no TMCI for positive modes.
	In both cases the lowest TMCI threshold is due to coupling of the
	$(-6)$-th and $(-7)$-th modes.
	}
\end{figure}

%------------------------------------------------------------------------------%
Figure~\ref{fig:SPS} shows the behavior of the lowest TMCI threshold as a 
function of space charge.
With the provided decay rate, the wake is effectively negative:
the only instabilities are in the negative part of spectrum and they all vanish
with the growth of the SC parameter.
A complementary Fig.~\ref{fig:SPSSpectra} shows the spectra for no space charge
case (left plot) and negative part of spectra when $\Qsc/\Qs = 20$.
The lowest TMCI threshold is associated with coupling of the $(-6)$-th and 
$(-7)$-th modes, following the resonator frequency $\omega_r$.  
For parameters from~\cite{quatraro2010effects}, the threshold number of 
particles at zero SC is $\Nb^\text{th} = 1.8\times 10^{11}$.

%------------------------------------------------------------------------------%
%------------------------------------------------------------------------------%

%------------------------------------------------------------------------------%
This result apparently contradicts~\cite{quatraro2010effects}, where the 
threshold at high value of SC parameter is almost the same as 
for the no space charge case. In their case, the space charge parameter, 
computed with the effective value of the space charge tune shift is $\Qsc/\Qs = 
23$; (the effective SC tune shift is a half of its maximum).
According to the Fig.~\ref{fig:SPS} the ratio of the thresholds with and 
without 
SC should be 
$\approx 3.5$.
The last is confirmed by all SSC models: there is no instability for 
effectively negative broadband wakes. 

To have more confidence in the ABS model, in its application to sinusoidal 
potential wells, we conducted its study for the high frequency broadband 
impedance and no SC, comparing its TMCI threshold with ones obtained by the NHT 
Vlasov solver \cite{PhysRevSTAB.17.021007} and pyHEADTAIL tracking simulations 
of A.~Oeftiger \cite{OeftigerPrivate}. As one may see from our 
Ref.~\cite{burov2018tmci}, there is a remarkably good agreement between all the 
three approaches in terms of the threshold values, although the numbers of the 
coupling modes are higher for the ABS model.   

As it was pointed out at the beginning of 
this article, for the vanishing type of TMCI, there is always an absolute 
threshold of the wake amplitude, such that the beam is stable for any number of 
particles as soon as the wake is below its absolute threshold value. Results of 
Fig.~\ref{fig:SPS} allow to compute the absolute threshold for the SPS 
parameters of Ref.~\cite{quatraro2010effects}, expressing it in terms of the 
shunt impedance. For transverse normalized rms emittance $\epsilon_n =1.5 
{\mathrm {mm\cdot mrad}}$ accepted at the referred paper, the absolute 
threshold 
comes out as $25 {\mathrm M\Omega/m}$. Thus, at this emittance, the accepted 
wake amplitude is about 5 times lower than the absolute threshold, and TMCI 
should not be seen with any number of particles. For higher emittance, the 
absolute threshold drops inversely proportionally to that.   

%------------------------------------------------------------------------------%

A resolution of the contradictions between the observations and macroparticle 
simulations for the SPS, on the one hand, and the Vlasov analysis of this and 
preceding publications, on the other, was recently suggested by one of the 
authors~\cite{burov2018convective}. As shown there, while SC moves up the TMCI 
wake threshold, it drives the {\it saturating convective 
instability} (SCI) and possibly the {\it absolute-convective instability} (ACI), 
pretty much 
for the same wake parameters, if not even lower.
As a result, the observed wake instability threshold can hardly be considerably 
larger than 
its no-SC value. 
\section{\label{sec:Summary}Summary}

%------------------------------------------------------------------------------%
We investigated transverse mode coupling thresholds for various 
space charge, wakes, potential wells and bunch distribution functions.
Two analytical approaches were used, the air-bag square well (ABS) model 
~\cite{blaskiewicz1998fast} and the strong space charge (SSC) theory 
~\cite{burov2009head,burov2015damping,balbekov2009transverse}. These approaches 
are complementary. The former is valid for a specific bunch configuration, 
allowing an arbitrary space charge, while the latter can be applied for any 
shape 
of the potential well and bunch 
distribution function, provided that SC is sufficiently strong.

As it was explained in Section~\ref{sec:Models}, there can be only two types of 
TMCI with respect to SC. 
The first one is {\it vanishing TMCI}: the threshold value of the wake tune 
shift $\kappa W_0$ grows linearly with the SC tune shift when the latter is 
high 
enough; this sort of instability cannot be 
seen in the SSC approximation, and that is why we call it vanishing. For the 
vanishing TMCI, there is an absolute threshold of the wake amplitude, such that 
the beam is stable for any number of particles, provided that the wake 
amplitude 
is below its absolute threshold value. Note that the latter is inversely 
proportional to the transverse emittance and beta-function.

The only 
other type of TMCI is the {\it SSC} one:
in this case the threshold of the wake tune shift is asymptotically inversely 
proportional
to the SC tune shift; in fact, it is proportional to $Q_s^2/\Qef(0)$. 
In our studies of particular examples, vanishing type of TMCI was always 
observed 
for negative and effectively negative wakes, which agrees 
with similar results of M.~Blaskiewicz and V.~Balbekov.
The same sort of TMCI has been found for resonator wakes with arbitrary decay 
rates, for all HP and Gaussian bunches inside a parabolic potential well.

The alternative type of TMCI, the SSC one, takes place for the cosine wakes 
without 
respect to the potential well and bunch distribution; it is also the case for 
the sine wakes at square potential wells, provided that phase advances of the 
oscillatory wakes are sufficiently large. 

We compared our results with available multiparticle simulations with space 
charge~\cite{
blaskiewicz2012comparing,quatraro2010effects}, and some contradictions were 
found. 
To convince ourselves of correctness of our results, we inspected the agreement 
of 
our two models with each other whenever possible; the numerical convergence 
was thoroughly examined always. For the same purpose, we solved Cauchy problem 
for the ABS model for several selected cases. For all of them, the thresholds 
obtained by the described eigensystem analysis were confirmed. A 
resolution of the contradiction was recently suggested by one of the authors in 
Ref.~\cite{burov2018convective}, where he treats the SPS Q26 instability not as 
TMCI but as a convective one.  

%To find 
%out causes of our contradictions with 
%Refs.~\cite{blaskiewicz2012comparing,quatraro2010effects}, various convergence 
%checks of the macroparticle simulations seem to be necessary. 

%==============================================================================%
%==============================================================================%
%==============================================================================%
\begin{acknowledgments}

%------------------------------------------------------------------------------%
Authors would like to thank Eric Stern and Stanislav (Stas) Baturin 
for their discussions and valuable input, and,
Francesca Weaver-Chaney and Lev Burov who assisted in the proof-reading of 
the manuscript.

%------------------------------------------------------------------------------%
This manuscript has been authored by Fermi Research Alliance, LLC under 
Contract 
No. DE-AC02-07CH11359 with the U.S. Department of Energy, Office of Science, 
Office of High Energy Physics. The U.S. Government retains and the publisher, 
by 
accepting the article for publication, acknowledges that the U.S. Government 
retains a non-exclusive, paid-up, irrevocable, world-wide license to publish or 
reproduce the published form of this manuscript, or allow others to do so, for 
U.S. Government purposes.
%The author would like to thank
%{\bf Leo~Michelotti},
%{\bf Eric~Stern}
%and
%{\bf James~F.~Amundson}
%for their discussions and valuable input.
%{\bf Alexey~Burov} for encouraging to find full family of solutions.
%{\bf Valeri~Lebedev} whose solution for electrostatic quadrupole led me to
%generalization, just as in the case with {\bf E.~M.~McMillan} and
%{\bf F.~Krienen}.
%And, of course, {\bf Sergei~Nagaitsev} who brought back to life original
%unknown McMillan's article which helped me with symmetric description of
%electromagnetic fields.
\end{acknowledgments}

%==============================================================================%
%==============================================================================%
%==============================================================================%
%==============================================================================%
%==============================================================================%
\appendix

%==============================================================================%
%------------------------------------------------------------------------------%
%------------------------------------------------------------------------------%
%------------------------------------------------------------------------------%
%==============================================================================%
\section{\label{secAP:Wlm}Matrix elements $\mW$ for SSC models}

%==============================================================================%
%------------------------------------------------------------------------------%
%==============================================================================%
\subsection{\label{secAP:WlmCONST}Analytical expressions}

%------------------------------------------------------------------------------%
Calculation of matrix elements for SSCSW model boils down to standard integrals 
of two trigonometric and an exponential (or inverse square root for resistive 
wall wake) functions.
They are easy to compute and the reader can find them using almost any symbolic 
computation program;
matrix elements for sine and cosine wake functions should be calculated with 
care when $\omega\,\tb = \pi\,n$ with $n$ being integer due to a resonance 
condition.

%------------------------------------------------------------------------------%
For the SSCHP$_0$ the matrix elements are not that trivial and we provide $\mW$
with the corresponding  wake function $W(\tau)$ in a Table~\ref{tab:WlmHP} 
below.

%==============================================================================%
%------------------------------------------------------------------------------%
%==============================================================================%
\subsection{\label{secAP:WlmNUMER}$\mW$ for SSCHP$_{1/2,1}$ and  SSCG models}

%------------------------------------------------------------------------------%
The numerical evaluation of the double integral of wake matrix elements
\[
\mW =	\int_{-\infty}^\infty \int_{\tau}^\infty
		W(\tau-\sigma)\,
		\rho(\tau) \, \rho(\sigma) \, Y_l(\tau) \, Y_m(\sigma)
		\,\dd\,\sigma\,\dd\,\tau
\]
is very expensive in terms of CPU time.
The use of bunch dipole moments
\begin{equation} 
I_{k}(\zeta) = \int_{-\infty}^{\infty} e^{i\,\zeta\,\tau}\,\rho(\tau)\,
Y_{k}(\tau)\,\dd\tau
\end{equation}
along with the impedance function
\begin{equation} 
	Z^\perp (\zeta) = i\,\int_{-\infty}^\infty 
W(\tau)\,e^{-i\,\zeta\,\tau}\,\dd\tau
\end{equation}
reduces the original expression to a single integral
\begin{equation}
\mW = (-1)^{m+1}\,i\,\int_{-\infty}^{\infty} Z^\perp(\zeta)\, 
	I_{l}(\zeta)\,I_{m}(\zeta)\,\frac{\dd\zeta}{2\,\pi}.
\end{equation}
Using this method, we are technically still calculating the double integral, 
but the integrals are decoupled and the inner integral ($I_k$) doesn't depend 
on a wake function and have to be found just once.

%==============================================================================%
%------------------------------------------------------------------------------%
%------------------------------------------------------------------------------%
%------------------------------------------------------------------------------%
%==============================================================================%
\section{\label{secAP:AB}Wakes for ABS model}

%------------------------------------------------------------------------------%
Matrix $\mathrm{M}$ for exponential, cosine and resonator wakes is provided
below in a Table~\ref{tab:MAB}.

%==============================================================================%
\begin{table*}[p!]
\caption{\label{tab:WlmHP}
	Matrix elements $\mW$ for SSCHP$_0$ model with negative (delta function,
	constant, exponential and resistive wall) and oscillating 
(trigonometric 
	and resonator) wakes.
	$\mW$ for all oscillating wakes is expressed using matrix elements of 
	the exponential one.
	}
\begin{ruledtabular}
\begin{tabular}{ll}
$W(\tau),\,\tau \leq 0$				& $\mW$		\\[-0.25cm]
								\\\hline
								\\[-0.1cm]
\multicolumn{2}{c}{Negative wakes}				\\[0.25cm]
$-W_0\,\delta(\tau)$				&
$\ds -W_0\,\delta_{l,m}$			\\[0.35cm]
$-W_0$						&
$\ds -\frac{W_0}{2}
	\left[
		\delta_{lm,l-m} -
	(-1)^{\left\lfloor{l/2}\right\rfloor + \left\lfloor{m/2}\right\rfloor}
		\frac{\delta_{l,m+1} - \delta_{l,m-1}}
		{\sqrt{(2\,l+1)(2\,m+1)}}
	\right]	$						\\[0.35cm]
$-W_0\exp(\alpha\,\tau)$			&
$\ds -W_0\sum_{lm}^{ij}
\frac{j!}{\alpha^{j+1}}
\sum_{k=0}^j 2^{j-k+1} \frac{\alpha^k}{k!}
\left\{
\frac{1+(-1)^{i+k}}{i+k+1} - \frac{e^{-\alpha/2}}{(-\alpha/2)^{i+1}}
\left[ \Gamma(i+1,\alpha/2) - \Gamma(i+1,-\alpha/2) \right]
\right\}$
\footnote{Where operator $
\sum_{lm}^{ij} =
(-1)^{\left\lfloor{l/2}\right\rfloor+\left\lfloor{m/2}\right\rfloor}\,
2^{l+m-2}\,\sqrt{(2\,l+1)(2\,m+1)}\,\sum_{i=0}^l\,\sum_{j=0}^m
	\begin{bmatrix}
		l		\\ i
	\end{bmatrix}
	\begin{bmatrix}
		m		\\ j
	\end{bmatrix}
	\begin{bmatrix}
		\frac{l+i-1}{2}	\\ l
	\end{bmatrix}
	\begin{bmatrix}
		\frac{m+j-1}{2}	\\ m
	\end{bmatrix}
$.}								\\[0.35cm]
$-W_0/\sqrt{|\tau|}$				&
$\ds -\frac{W_0}{\sqrt{2}}
	\sum_{lm}^{ij}\frac{2^{i+j+\frac{3}{2}}}{i+j+\frac{3}{2}}\,
	\sum_{k=0}^j \frac{(-1)^{i+k}}{j-k+\frac{1}{2}}\,
	\begin{bmatrix}
	j \\ k
	\end{bmatrix}\,
	_{2}F_{1}\left[
		-\left( i+j+\frac{3}{2} \right);
		-\left( i+k             \right);
		-\left( i+j+\frac{1}{2} \right);
		\frac{1}{2}
	\right]$						\\[0.65cm]
\multicolumn{2}{c}{Oscillating wakes}				\\[0.25cm]
$-W_0\cos(\omega\,\tau)$			&
$ \Re\left[\mW^\mathrm{exp}(i\,\omega)\right]$			\\[0.45cm]
$ W_0\sin(\omega\,\tau)$			&
$-\Im\left[\mW^\mathrm{exp}(i\,\omega)\right]$			\\[0.45cm]
$ W_0\sin(\omega\,\tau)\,\exp(\alpha\,\tau)$	&
$-\Im\left[\mW^\mathrm{exp}(\alpha+i\,\omega)\right]$
\end{tabular}
\end{ruledtabular}
\end{table*}
%==============================================================================%

%==============================================================================%
\begin{table*}[h]
\caption{\label{tab:MAB}
	Matrix $\mathrm{M}$ for ABS model with exponential (or constant for 
	 $\alpha = 0$), cosine and resonator (or sine for $\alpha=0$) wake 
	functions.
	}
\begin{ruledtabular}
\begin{tabular}{lc}
$W(\tau),\,\tau\leq 0$	& $\mathrm{M}$	\\[-0.25cm]
				\\\hline
				\\[-0.1cm]
$-W_0\,e^{\alpha\,\tau}$				&
$
\begin{bmatrix}
 i\,\pi\left( \frac{1}{2}\frac{\Qsc}{\Qs} + \frac{\Qx}{\Qs} \right)	&
-i\,\pi\,\frac{1}{2}\frac{\Qsc}{\Qs}					&
 i\,\pi									\\
 i\,\pi\,\frac{1}{2}\frac{\Qsc}{\Qs}					&
-i\,\pi\left( \frac{1}{2}\frac{\Qsc}{\Qs} + \frac{\Qx}{\Qs} \right)	&
-i\,\pi									\\
-\frac{1}{2}\frac{\kappa\,W_0\tb}{\Qs}				&
-\frac{1}{2}\frac{\kappa\,W_0\tb}{\Qs}				&
 \alpha\,\tb								\\
\end{bmatrix}
$							\\[1.cm]
$-W_0\,\cos(\omega\,\tau)$				&
$
\begin{bmatrix}
 i\,\pi\left( \frac{1}{2}\frac{\Qsc}{\Qs} + \frac{\Qx}{\Qs} \right)	&
-i\,\pi\,\frac{1}{2}\frac{\Qsc}{\Qs}					&
 i\,\pi									&
 i\,\pi									\\
 i\,\pi\,\frac{1}{2}\frac{\Qsc}{\Qs}					&
-i\,\pi\left( \frac{1}{2}\frac{\Qsc}{\Qs} + \frac{\Qx}{\Qs} \right)	&
-i\,\pi									&
-i\,\pi									\\
-\frac{1}{4}\frac{\kappa\,W_0\tb}{\Qs}				&
-\frac{1}{4}\frac{\kappa\,W_0\tb}{\Qs}				&
 i\,\omega\,\tb								&
 0									\\
-\frac{1}{4}\frac{\kappa\,W_0\tb}{\Qs}				&
-\frac{1}{4}\frac{\kappa\,W_0\tb}{\Qs}				&
 0									&
-i\,\omega\,\tb
\end{bmatrix}
$							\\[1.cm]
$ W_0\,\sin(\omega\,\tau)\,e^{\alpha\,\tau}$		&
$
\begin{bmatrix}
 i\,\pi\left( \frac{1}{2}\frac{\Qsc}{\Qs} + \frac{\Qx}{\Qs} \right)	&
-i\,\pi\,\frac{1}{2}\frac{\Qsc}{\Qs}					&
 i\,\pi									&
 i\,\pi									\\
 i\,\pi\,\frac{1}{2}\frac{\Qsc}{\Qs}					&
-i\,\pi\left( \frac{1}{2}\frac{\Qsc}{\Qs} + \frac{\Qx}{\Qs} \right)	&
-i\,\pi									&
-i\,\pi									\\
-\frac{i}{4}\frac{\kappa\,W_0\tb}{\Qs}				&
-\frac{i}{4}\frac{\kappa\,W_0\tb}{\Qs}				&
 (\alpha+i\,\omega)\,\tb							
	&
 0									\\
 \frac{i}{4}\frac{\kappa\,W_0\tb}{\Qs}				&
 \frac{i}{4}\frac{\kappa\,W_0\tb}{\Qs}				&
 0									&
 (\alpha-i\,\omega)\,\tb
\end{bmatrix}
$
\end{tabular}
\end{ruledtabular}
\end{table*}
%==============================================================================%

%==============================================================================%
%------------------------------------------------------------------------------%
%------------------------------------------------------------------------------%
%------------------------------------------------------------------------------%
%==============================================================================%
\section{\label{secAP:Spectra}Additional spectra for SSC models}

\begin{figure*}[th!]
\includegraphics[width=\linewidth]{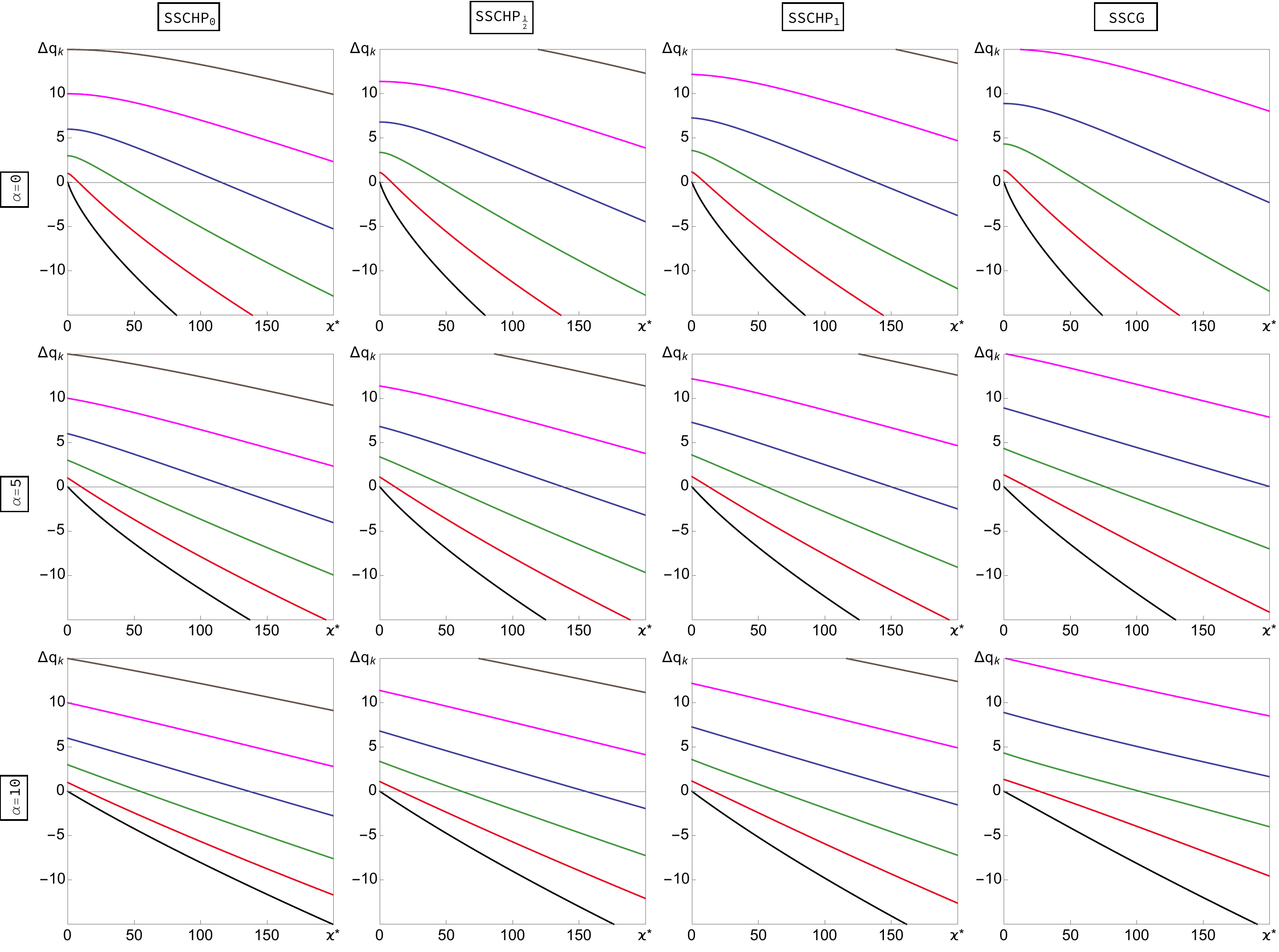}
\caption{\label{fig:SSCExpWakeSpectraAPPNDX}
	Bunch spectra for SSCHP$_{0,\frac{1}{2},1}$ (1st -- 3rd 
	columns) and SSCG (4th column) models with the constant and
	exponential wakes.
	Different rows correspond to the different values of
	$\alpha = 0,5,10$ measured in units of $\tb$ with $\tb = 3\,
	\cb$ for SSCG.
	}
\end{figure*}

\begin{figure*}[th!]
\includegraphics[width=\linewidth]{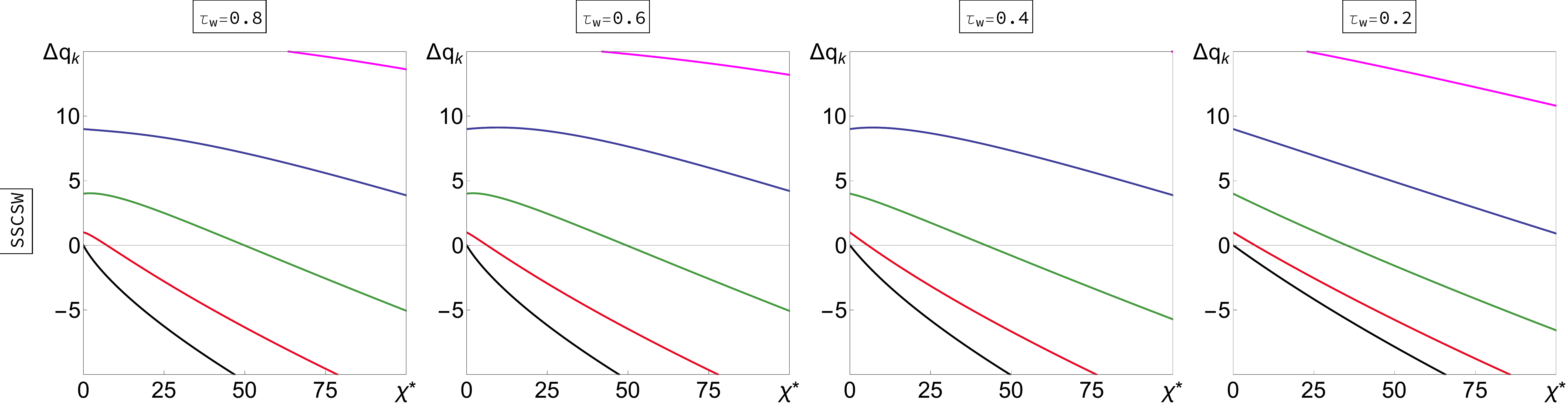}
\includegraphics[width=\linewidth]{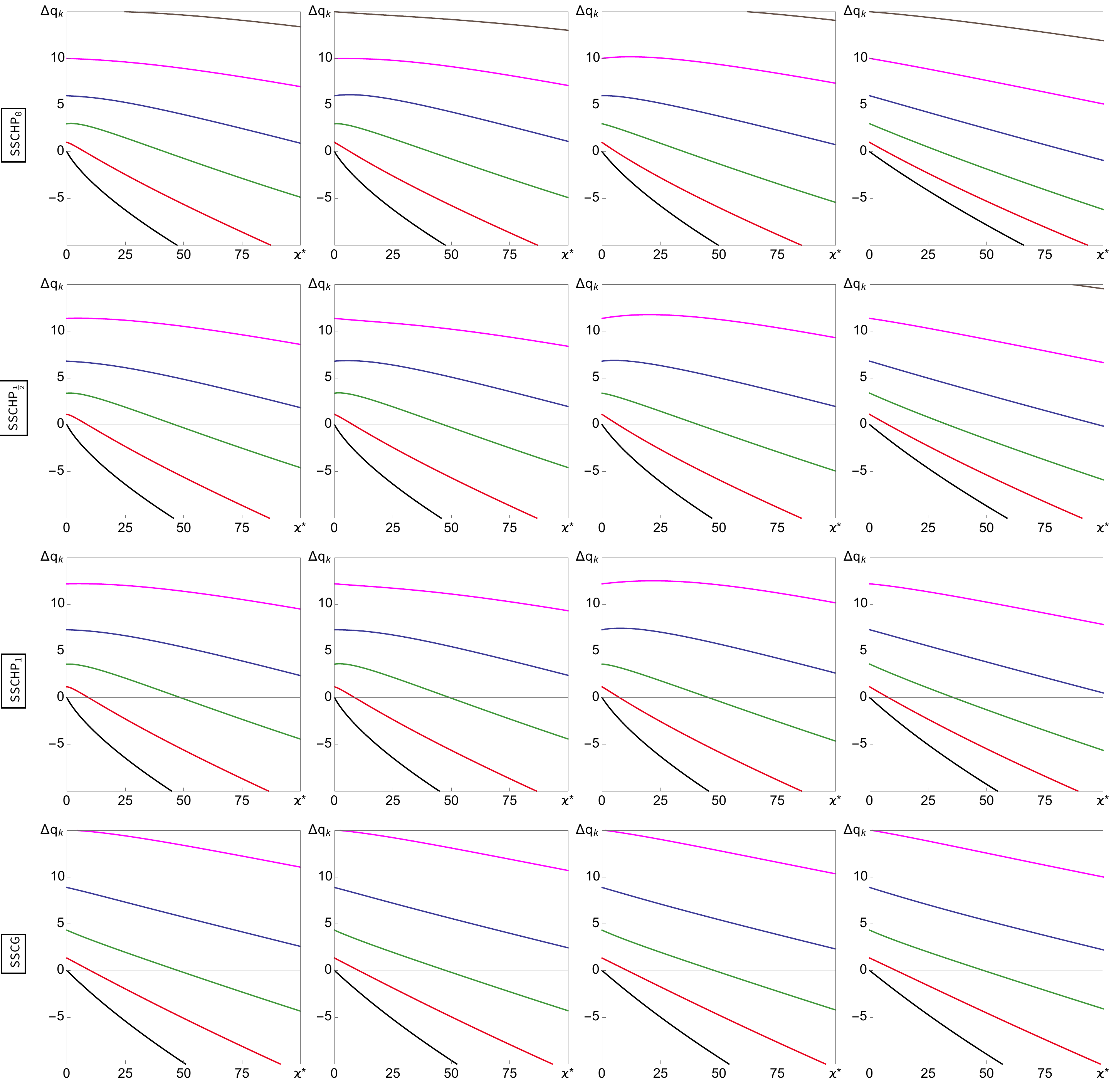}
\caption{\label{fig:SSCStepWakeSpectraAPPNDX}
	Bunch spectra for all SSC models with the step wake.
	Different columns correspond to the different values of wake length
	$\tau_\text{w} = 0.8,0.6,0.4$ and $0.2$ measured in units of $\tb$ (or 
$3\,\cb$
	for SSCG).
	}
\end{figure*}

\begin{figure*}[th!]
\includegraphics[width=\linewidth]{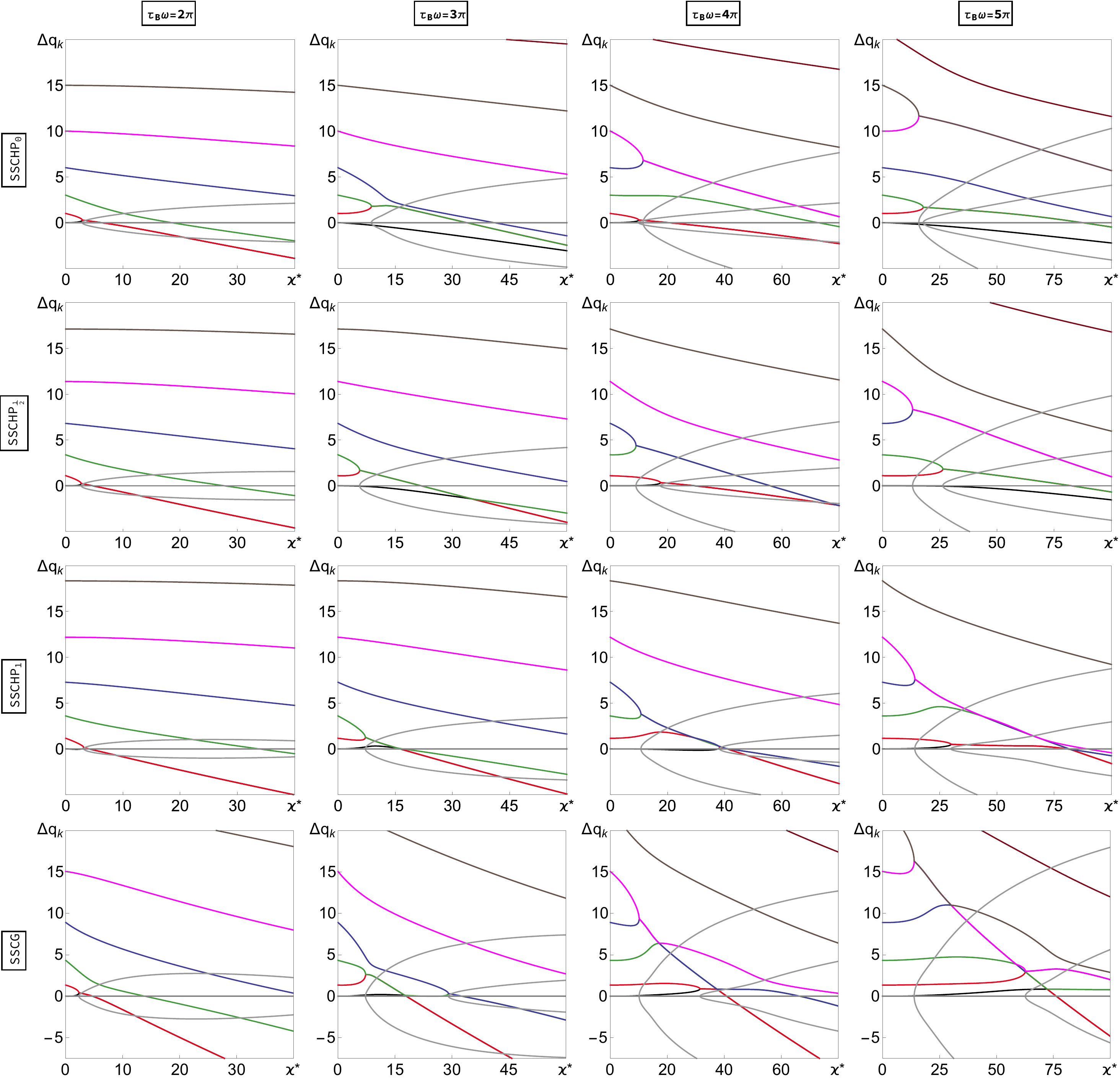}
\caption{\label{fig:SSCCosWakeSpectraAPPNDX}
	Bunch spectra for SSCHP$_{0,1/2,1}$ (1st -- 3rd rows) and SSCG 
	(bottom row) models with the cosine wake.
	Real and imaginary parts of the spectra are shown in colors and gray 
	respectively.
	The columns correspond to the different values of wake phase advance 
	$\omega\,\tb$ with $\tb = 3\,\cb$ for SSCG.
	}
\end{figure*}

%------------------------------------------------------------------------------%
Complementary SSC bunch spectra for constant and exponential
(Fig.~\ref{fig:SSCExpWakeSpectraAPPNDX}),
step-function
(Fig.~\ref{fig:SSCStepWakeSpectraAPPNDX})
and cosine
(Fig.~\ref{fig:SSCCosWakeSpectraAPPNDX})
wakes are provided below.
%------------------------------------------------------------------------------%
%------------------------------------------------------------------------------%

%------------------------------------------------------------------------------%
%------------------------------------------------------------------------------%

%------------------------------------------------------------------------------%
%------------------------------------------------------------------------------%

%\nocite{*}

%\bibliography{bibfile}			% Produces the bibliography via BibTeX.

\providecommand{\noopsort}[1]{}\providecommand{\singleletter}[1]{#1}%

\end{document}